\documentclass[%
aps,
reprint,
nofootinbib,
amsmath,amssymb,
pra,
]{revtex4-2}

\PassOptionsToPackage{color}{xy}
\usepackage[usenames,dvipsnames]{xcolor}

\usepackage{mathrsfs}
\usepackage{dsfont}
\usepackage{bbm}
\usepackage{palatino}

\usepackage[hmargin=2cm,vmargin=2cm]{geometry}

\usepackage{titlesec} 
\titlespacing*{\paragraph}{\parindent}{1.8ex plus .2ex minus .2ex}{2ex plus .5ex minus .2ex}

\usepackage{enumitem}

\usepackage{calc}     
\usepackage{tikz}     
\usepackage{graphicx} 
\graphicspath{{./figures/}}
\usepackage{float}
\usepackage[font=small]{caption}
\usepackage{subcaption}
\usepackage{ragged2e}
\DeclareCaptionJustification{justified}{\justifying}
\makeatletter
\renewcommand{\fnum@figure}{Fig. \thefigure}
\makeatother

\usepackage{physics}
\usepackage[utf8]{inputenc}    
\usepackage[english]{babel}    
\usepackage{array}    
\usepackage{braket}   
\usepackage{xfrac}    
\usepackage{units}    
\usepackage[colorlinks=true,urlcolor=black,linkcolor=Mahogany,citecolor=Mahogany,plainpages=false,pdfpagelabels]{hyperref}
\usepackage{url}

\newcommand{\wqbit}{\omega_\text{L}}
\newcommand{\wqbitk}[1]{\omega_{\text{L},#1}}
\newcommand{\wdelta}{\delta\omega_\text{L}}
\newcommand{\wadd}{ \omega_\text{add}}

\newcommand{\wref}{\omega_\text{r}}
\newcommand{\wdeltak}[1]{\delta\omega_{\text{L},#1}}
\newcommand{\wrefk}[1]{\omega_{\text{r},#1}}
\newcommand{\waddk}[1]{ \omega_{\text{add},#1}}

\newcommand{\tpulse}{t_{\text{pulse}}}

\newcommand{\DEZ}{\Delta E_{rZ}}

\newcommand{\VC}{V_\text{C}}
\newcommand{\VE}{V_\text{E}}
\newcommand{\VEkk}[1]{V_{\text{E},#1}}
\newcommand{\TL}{T_{2,L}^*}

\newcommand{\tQ}{t_{\text{\sc qec}}}

\newcommand{\nQD}{n_{\text{\sc qd}}}

\newcommand{\ie}{\emph{i.e.}}
\newcommand{\eg}{\emph{e.g.}}

\begin{document}

\title{Simulated non-Markovian Noise Resilience of Silicon-Based Spin Qubits\\ with Surface Code Error Correction}

\author{Oscar Gravier$^{(1)}$, \quad Thomas Ayral$^{(2)}$, \quad Beno\^it Vermersch$^{(3)}$, \quad Tristan Meunier$^{(3,4)}$, \quad Valentin Savin$^{(1)}$\\[2mm]
$^{(1)}$\,Universit\'e Grenoble Alpes, CEA-L\'eti, F-38054 Grenoble, France \\
$^{(2)}$\,Eviden Quantum Laboratory, F-78340 Les Clayes-sous-Bois, France \\
$^{(3)}$\,Quobly Grenoble, France
$^{(4)}$\,Institut Néel, CNRS Grenoble, France}

\begin{abstract}
We investigate the resilience of silicon-based spin qubits against non-Markovian noise within the framework of quantum error correction. We consider a realistic non-Markovian noise model that affects both the Larmor frequency and exchange energy of qubits, allowing accurate simulations of noisy quantum circuits. We employ numerical emulation to assess the performance of the \mbox{distance-3} rotated surface code and its XZZX variant, using a logical qubit coherence time metric based on Ramsey-like experiments. Our numerical results suggest that quantum error correction converts non-Markovian physical noise into Markovian logical noise, resulting in a quartic dependence of coherence time between physical and logical qubits. Additionally, we analyze the effects of spatial noise correlations and sparse architectures, substantiating the robustness of quantum error correction in silicon-based spin qubit systems.
\end{abstract}

\maketitle

Quantum computers hold the promise of increased computational power in many areas, including factoring \cite{miquel_factoring_1996}, optimization  \cite{farhi_quantum_2001}, quantum simulation \cite{Lloyd1996} and quantum chemistry \cite{aspuru-guzik_simulated_2005}. 
However, achieving  quantum advantage in these fields requires computers that operate with extremely low error rates due to the exponential decrease in fidelity with the number of operations~\cite{louvet2023feasibility}. Achieving such low error rates remains beyond the reach of current technology, as quantum devices are inherently prone to noise, and whether such performance levels are ultimately attainable remains an open question. To circumvent this exponential wall, quantum error correction (QEC) has been proposed early in~\cite{Shor1995}.
This technique protects the quantum information by encoding it across multiple physical qubits.
It enables exponential error suppression by scaling the number of physical qubits, provided  the physical error rate remains below a certain threshold.
With recent demonstrations of small QEC codes on a variety of hardware platforms, including superconducting qubits~\cite{kim2023evidence, acharya_quantum_2024}, trapped ions~\cite{valentini2024demonstration}, and Rydberg atoms~\cite{radnaev2024universal, reichardt2024logical}, accurate modeling and prediction of QEC experiments via classical numerical simulations becomes a pressing issue. Most such simulations assume Clifford-only operations and Pauli-only noise models to achieve efficient classical simulations~\cite{gottesman1997stabilizer, aaronson2004improved, higgott2025sparse}. However, more realistic (non-Clifford or non-Pauli) models (see \eg~\cite{darmawan2017tensor, suzuki2017efficient}) are likely necessary to inform the scalable development of near-term quantum devices with integrated error correction.

In this work, we focus on the silicon-based spin qubit technology, which benefits from advanced manufacturing techniques established in the semiconductor industry, promising significant scalability for hardware devices~\cite{burkard_semiconductor_2023}. 
Our aim is to evaluate the performance of surface codes \cite{fowler_surface_2012} on this hardware, with realistic assumptions as to the underlying noise models. In an effort to comprehensively capture the hardware characteristics, we employ realistic correlated noise models and native (non-Clifford) operations, using efficient state-vector emulation to assess the performance of small QEC codes on this hardware.

QEC on  semiconductor spin qubits has recently been  studied in~\cite{hetenyi_tailoring_2024, pataki_coherent_2024}.
Ref.~\cite{hetenyi_tailoring_2024} focuses on  adapting  the surface code circuit to the characteristic features of spin qubits (\eg, in case  an additional qubit is needed for readout,  necessitating careful adjustments of the qubit layout), but the error correction performance is assessed under simplified Pauli noise models. Ref.~\cite{pataki_coherent_2024} considers a quasistatic phase damping error model,  consisting
of coherent rotations around the $Z$-axis for each data qubit, with independent and identically distributed (i.i.d) rotation angles drawn from a  Gaussian distribution with zero mean. Syndrome measurement circuits are assumed perfect, but measurement outcomes are flipped with some non-zero probability.

Here, we consider a realistic error model that accounts for non-Markovian noise affecting both the Larmor frequency and the exchange energy of the qubits~\cite{dial2013charge}.
Our error model captures not only the decoherence of idling qubits, but also the noise in Si-qubit native gates operating on one and two qubits, enabling faithful simulations of noisy  circuits. We consider the distance-3 rotated surface code (and its XZZX variant~\cite{bonilla_ataides_xzzx_2021}) and assess the performance of the logical qubit, based on the emulation of noisy syndrome measurement circuits  on the Qaptiva\textsuperscript{\sc tm} quantum emulator.
We define the logical qubit coherence time through a Ramsey-like experiment on the logical qubit, and use it as a performance metric directly comparable to the coherence time of the physical qubit.
Our numerical results suggest that quantum error correction  \emph{Markovianizes} noise, that is, it converts non-Markovian physical noise into Markovian logical noise. As a result, the quadratic dependence between the error rates of logical and physical qubits shifts to a quartic dependence in terms of coherence time, leading to a substantial improvement in the logical qubit coherence time compared to the physical qubit.
Finally, we consider two scenarios practically relevant for real-world systems: we include (i) spatial noise correlations originating from the shared environment  between qubits, and (ii) sparse architectures that may be needed, for instance, to simplify wiring or to accommodate additional qubits for spin-to-charge conversion-based readout.    
Our results show that the surface code maintains strong error suppression capabilities in these two scenarios.

The article is organized as follows. Section~\ref{sec:hardware_noise_model} presents the hardware  model, including our  main working assumptions and the Hamiltonians of one and two-qubit gates. Native gates are then discussed in Section~\ref{sec:native-gates}, along with their  noise model and the underlying non-Markovian process.  Section~\ref{sec:qec_emulation} presents the surface code emulation for spin-based qubits, including the details of the spatially correlated noise and sparse architecture scenarios. Section~\ref{sec:performance-metrics} introduces the characteristic time  metrics used to evaluate the performance of both physical and logical qubits, and numerical results are presented and discussed in Section~\ref{sec:numerical-results}.

\section{Microscopic Hardware Model}
\label{sec:hardware_noise_model}

\begin{figure}[!t]
    \centering
    \includegraphics[width=\linewidth]{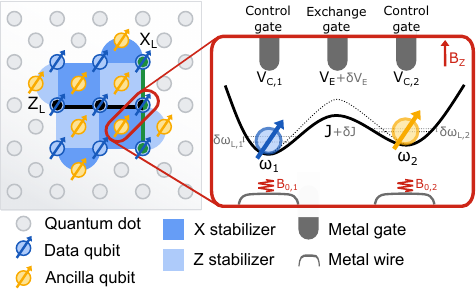}
    \caption{ (left) A 2D lattice of quantum dots/spin qubits. A distance-3 rotated surface code (17 qubits) is also shown, with data qubits depicted in blue, and ancilla qubits depicted in yellow. Dark  and light blue patches correspond to $X$ and $Z$ stabilizer generators, respectively.  
    (right) Electromagnetic environment of two qubits. Relevant parameters and hardware elements discussed in the text are also shown.}
    \label{fig:rotated_surface17}
\end{figure}

We consider a 2D lattice of spin qubits on the surface of a silicon layer, as depicted in Fig.~\ref{fig:rotated_surface17}. Each qubit is made of a single electron trapped in a quantum dot (QD) and can interact with its four closest neighbors (see~\cite{burkard_semiconductor_2023} for a review on silicon-based quantum technology). The states of the qubits are manipulated by changing the trapping potential through the electrical potential of control gates ($\VC$) and exchange gates ($\VE$), or by using electron spin resonance (ESR) with a driving field ($B_0$) generated by a metal wire \cite{koppens_koppens_2006, nowack_coherent_2007}. Various size-restricted structures have been developed and are presented in~\cite{ mortemousque_coherent_2021, hendrickx_four-qubit_2021}, and a good introduction on how to fill a 2D structure with electrons can be found in~\cite{gonzalez-zalba_scaling_2021}.

During state manipulation, a   magnetic field $B_Z$ is applied, inducing a precession of the spin about the $Z$-axis, with an angular frequency known as Larmor frequency, and denoted by $\wqbit$. The Larmor frequency must be continuously tracked relative to a reference frequency $\wref$ throughout the computation. Each qubit has its own individual reference frequency, allowing us to abstract away their individual phase evolution. 

In such a device, both the Larmor frequency and exchange gate potential are subject to non-Markovian noise, leading to  deviations $\wdelta$ and $\delta \VE$ from their desired values. Non-Markovian noise refers to the fact that it is not possible to model the stochastic Hamiltonian evolution via a Lindblad equation, and in our case this originates from the time correlations present in the Hamiltonian parameters. In contrast with white noise models, one obtains a dephasing master equation that corresponds to a Pauli channel~\cite{preskill1998lecture}. 
The $\delta \VE$ deviation also causes a deviation $\delta J$ of the coupling parameter $J$. The non-Markovian characteristic of the noise, as well as the relation between  $\delta \VE$ and $\delta J$ deviations are discussed in Section~\ref{subsec:non-markovian-noise}. We assume that, except for two-qubit gate manipulations, each exchange gate potential can be set to a position that enables zero $J$ coupling between the qubits.

We provide below a brief description of the Hamiltonians corresponding to single and two-qubit gates.

\subsection{Single-Qubit Hamiltonian}
To simulate single-qubit gates, we use the following Hamiltonian\footnote{To align with experimental data, all Hamiltonian operators are expressed in units of frequency (Planck constant $h = 1$).}, written in a frame rotating with a reference Larmor frequency  $\wref$, %
\begin{equation}
    \hspace*{-2mm}\resizebox{.93\linewidth}{!}{\( \displaystyle H(t) = \frac{\wdelta(t) + \wadd(t)}{2}Z
    +
    \frac{B_0(t)}{2}\left(\cos(\varphi) X + \sin(\varphi) Y\right),\)\!\!} \label{eq:1q_H}
\end{equation}
where $\wadd(t)$ is a  controlled shift of the reference Larmor frequency,  $\wdelta(t) := \wqbit(t) -  \wref $ is the deviation of the qubit Larmor frequency from its reference, and $B_0(t)$ and $\varphi$ denote respectively the amplitude and the phase of the driving field.
 Note that we shall consider $\wadd(t) = 0$, except for gates without driving field, that is, for which $B_0(t)=0$. 
The Larmor frequency deviation  $\wdelta(t)$ represents the noise term in the expression of the Hamiltonian. In the noiseless case (\ie, $\wdelta=0$), appropriate choices of the driving field and Larmor frequency shift parameters in~(\ref{eq:1q_H}) lead to the implementation of the following native gates: Pauli matrices $X$, $Y$, $Z$, as well as,
\begin{equation}
    S :=
    \begin{bmatrix}
        1 & 0 \\
        0 & i
    \end{bmatrix}, \quad
    K_\varphi = \frac{1}{\sqrt{2}}
    \begin{bmatrix}
    1                            & -ie^{-i \varphi} \\
    -ie^{i \varphi}    & 1
    \end{bmatrix},
    \label{eq:S-Kphi-gate}
\end{equation}  as detailed in Section~\ref{sec:one_qubit_gate}.
The gate $K_\varphi$ corresponds to a rotation by an angle of $\pi/2$ about the \mbox{$\vec{r}(\varphi) =(\cos\varphi, \sin\varphi, 0)$} axis on the Bloch sphere. 
Finally, we  note that idle noise resulting from the deviation $\wdelta$ corresponds to $\wadd(t)=B_0(t)=0$ in~(\ref{eq:1q_H}), yielding a qubit  rotation about the $Z$-axis.

\subsection{Two-Qubit Hamiltonian}

For the two-qubit gate, we use subscript $k=1,2$ to indicate the corresponding qubit, and set $\waddk{k} = 0$. Thus, in the frame rotating with the Larmor frequency of qubit $1$, $\wrefk{1}$, the two-qubit Hamiltonian reads~\cite{burkard_semiconductor_2023},
\begin{multline}
    H_{r1}(t) =\frac{J(t)  + \delta J(t)}{4}\left(XX+YY+ZZ\right) \\
    +\frac{\wdeltak{1}(t)}{2}ZI+\frac{(\wdeltak{2}(t)+\DEZ)}{2}IZ, \label{eq:2q_H}
\end{multline}
where $\wdeltak{k}(t) := \wqbitk{k}(t) - \wrefk{k}$ is the deviation of the qubit Larmor frequency from its reference, $\DEZ := \wrefk{2} - \wrefk{1}$ is called the energy difference, and $J(t)$ is the coupling parameter. Noise is modeled here though deviation parameters $\wdeltak{1}(t), \wdeltak{2}(t)$ and $\delta J(t)$.  Section~\ref{sec:two_qubit_gate} describes the implementation and simulation of the  following two-qubit gate:
\begin{equation}
    P:=\left[\begin{array}{cccc}1 & 0 & 0 & 0 \\ 0 & i & 0 & 0 \\ 0 & 0 & i & 0 \\ 0 & 0 & 0 & 1 \end{array}\right] = CZ(S\otimes S), \label{eq:P_gate_def}
\end{equation}
where $CZ$ denotes the controlled-$Z$ gate.

\section{Native Gates Implementation}
\label{sec:native-gates}
In this section, we discuss the implementation of single-qubit (Section~\ref{sec:one_qubit_gate}) and two-qubit (Section~\ref{sec:two_qubit_gate}) native gates using time dependent controls on the Hamiltonians introduced in the previous section. Finally, Section~\ref{subsec:non-markovian-noise} describes the non-Markovian noise process studied and its impact on qubit manipulation parameters.

\subsection{Single-Qubit Natives Gates}\label{sec:one_qubit_gate}

We consider two different types of single-qubit gates.  For rotations about the $X$ or \mbox{$Y$-axes}, we use an oscillating magnetic  field $B_0(t)$ (Fig.~\ref{fig:rotated_surface17}) to drive the qubit by producing Rabi oscillations. For rotations about the \mbox{$Z$- axis}, we use the control gates in Fig.~\ref{fig:rotated_surface17} to induce a controlled shift $\wadd(t)$ of the reference Larmor frequency, thus generating a phase shift. For $B_0(t)$, we use a Gaussian pulse, which has been observed to yield better performance than square or cosine pulses in terms of gate fidelity (see Appendix~\ref{ap:pulse_shape_study}). In contrast, for $\wadd(t)$ we use a cosine pulse,  due to its  comparable performance and the advantage of enabling shorter gate times.

\subsubsection{Gates with driving field ($X$,$Y$ and $K$-family)}
\label{subsec:gates-with-driving-field}

For theses gates, we use a driving field $B_0(t)$ and set $\wadd(t)=0$.
We consider first the case of ideal gates ($\wdelta=0$), for which the  Hamiltonian~(\ref{eq:1q_H}) rewrites as 
$H(t) = \frac{1}{2}B_0(t)\left(\cos(\varphi) X + \sin(\varphi) Y\right)$. Since $H$  is time-commuting (that is, $H(t_1) H(t_2) = H(t_2) H(t_1)$, $\forall t_1, t_2$), the  corresponding  time evolution operator is given by:
 \begin{align}
  \hspace{-2mm}  U_\varphi(t) & = \exp \left( -i 2\pi \int_0^t H(s)ds  \right)                                                               \\
  & = \exp\left( -i \frac{\theta(t)}{2} (\cos(\varphi) X + \sin(\varphi) Y) \right) \\
                 & \quad \text{where } \theta(t) = 2\pi \int_0^t B_0(s) ds, \label{eq:thetaB}
\end{align}
corresponding to a rotation by an angle $\theta(t)$ about the $\vec{r}(\varphi) = (\cos(\varphi), \sin(\varphi), 0)$ axis on the Bloch sphere. 

For the driving field, we consider an oscillating electromagnetic field with Gaussian pulse shape
\begin{equation}\label{eq:gaussian-pulse}
B_0(t) = B_0 \, \exp\left(- \displaystyle \frac{(t-\tpulse/2)^2}{2\sigma^2}\right),
\end{equation}
which is applied for a duration $\tpulse$,  where $B_0$ is the amplitude of the pulse,  and $\sigma$ is the standard deviation controlling the pulse width. From~(\ref{eq:thetaB}), we get
\begin{equation}
  \theta(\tpulse) = 2\pi\, \sqrt{2\pi} B_0 \sigma \erf\left( \frac{\tpulse}{2\sqrt{2}\sigma} \right), \label{eq:thetaB-tpulse}
  \end{equation}
 where $\erf(x) := \frac{2}{\sqrt{\pi}}\int_0^x e^{-s^2} ds$ is the error function. We further define $b := B_0(0) = B_0(\tpulse) = B_0 e^{ - \tpulse^2 / (8\sigma^2) } $,  the value of the pulse at its limits. For given $b$ and $B_0$ values, a gate achieving a desired angle of rotation $\theta(\tpulse)$  can be implemented by taking
 \begin{align}
 \sigma &= \frac{\theta(\tpulse)}{(2\pi)^{\frac{3}{2}} \, B_0 \, \erf\left( \sqrt{\ln(B_0/b)} \right)}, \label{eq:sigma-derivation}\\
 \tpulse &=  2\sigma \sqrt{2\ln(B_0/b)}. \label{eq:tpulse-derivation}
 \end{align}
In our simulations, we consider Gaussian pulses with $b=1$\,Hz, and the $B_0$ value is chosen to align with ESR qubit manipulation (see Section~\ref{subsec:simulation-parameters}). We consider the following two families of gates, corresponding to either  $\pi/2$ or  $\pi$-rotations  on the Bloch sphere. We define these gates by specifying the $\theta(\tpulse)$ value, since the corresponding $\sigma$ and $\tpulse$ values can then be determined   from~(\ref{eq:sigma-derivation}) and~(\ref{eq:tpulse-derivation}).

\paragraph{The $K$-family gates:} For $\theta(\tpulse) = \displaystyle {\pi}/{2}$, we denote the corresponding  time evolution operator by $K_\varphi := U_{\varphi} (\tpulse)$, whose matrix expression is given in~(\ref{eq:S-Kphi-gate}). When the phase $\varphi$ is a multiple of $\pi/2$, we further define $K_n := K_{n {\pi}/{2}}$, corresponding to a rotation by an angle of $\pi/2$ about the $\vec{r}(n  \frac{\pi}{2})$ axis on the Bloch sphere.
Note that $K_1 = HZ = XH$, where $H$ denotes the Hadamard gate, and we also have $K_1^\dagger = K_3$ and $K_2 = K_0^\dagger$.

\paragraph{The $\pi$-pulse-family gates:} For  $\theta(\tpulse) = \pi$, and phase $\varphi = 0$ or $\varphi = \pi/2$, the above time evolution operator corresponds to either $X$ or $Y$ Pauli operator (up to a global phase), precisely $U_0(\tpulse) = -iX$,  and $U_{\pi/2}(\tpulse) = -iY$.

\medskip Finally, we note that for noisy gates (that is, when $\wdelta \neq 0$), the gate Hamiltonian is not time-commuting and thus the time evolution operator cannot be determined analytically. However it can be approximated numerically,  by using Trotter–Suzuki formulas~\cite{suzuki1992general}.

\subsubsection{Gates without driving field ($Z$-axis rotations)}
\label{subsec:gates-wout-driving-field}

To implement rotations about the $Z$-axis, we use no driving field ($B_0(t)=0$), but instead use a controlled shift, $\wadd(t)$, of the reference Larmor frequency.  

We consider first the case of ideal gates \mbox{($\wdelta=0$)}, for which the gate Hamiltonian~(\ref{eq:1q_H}) rewrites as \mbox{$H(t) = \frac{1}{2}\wadd(t)Z$}, and the corresponding  time evolution operator is given by:
\begin{align}
  \hspace*{-1mm}  R_Z\big( \zeta(t) \big) &:=
   \exp\left(\! -i\frac{\zeta(t)}{2}Z\right) \label{eq:unit-evolution-Z}\\                                      
     & \quad \ \text{where } \zeta(t) = 2 \pi \int_0^t  \wadd(s)\, ds, \label{eq:zeta}
\end{align}
corresponding to a rotation by an angle $\zeta(t)$ about the $Z$-axis on the Bloch sphere.

For $\wadd(t)$, we consider a cosine pulse 
\begin{align} \label{eq:cosine-pulse-omega}
\wadd(t) &= \displaystyle \frac{\omega_{0}}{2} \left(1-\cos(2\pi \frac{t}{\tpulse})\right), 
\end{align}
which is applied for a duration $\tpulse$ (note that we use the same notation, $\tpulse$, for the duration of different pulses, with the specific pulse referred to being clear from the context).  From~(\ref{eq:zeta}), we get
\begin{equation}
\zeta(\tpulse) = \omega_0\, \tpulse\, \pi.
\end{equation}

Thus, for $\omega_0$ and $\tpulse$ values  such that $\zeta(\tpulse) = \pi$ or $\zeta(\tpulse) = \pm\pi/2$, one obtains either the Pauli  $Z$ gate or the phase gates $S$ and $S^\dagger$ (up to global phases).

\medskip For noisy gates, the gate Hamiltonian is given by
\mbox{$H(t) = \frac{1}{2}(\wdelta(t)+\wadd(t))Z$}. The unitary time evolution operator takes the same form as in~(\ref{eq:unit-evolution-Z}), with rotation angle   \mbox{$\zeta(t) = 2 \pi \int_0^t \wdelta(s) + \wadd(s)\, ds$},  hence inducing an extra rotation by an angle of  \mbox{$2 \pi \int_0^t \wdelta(s) \,ds$}  compared to the ideal gate.
 Similarly,  qubits that remain idle during some time period $t$ undergo dephasing throughout that duration, inducing a rotation by an angle $2 \pi \int_0^t \wdelta(s)\, ds$ about the $Z$-axis, which we shall refer to as idle noise.

\subsection{Two-Qubit Native Gate}\label{sec:two_qubit_gate}

We consider the two-qubit Hamiltonian $H_{r1}$~(\ref{eq:2q_H}), in the reference frame of qubit~1,  with coupling parameter
\begin{equation}
    J(t) = \displaystyle\frac{J_0}{2} \left( 1-\cos\left(2\pi \frac{t}{\tpulse}\right)  \right), \label{eq:J-pulse-def}
\end{equation} which is applied for a duration $\tpulse$. The corresponding time evolution operator is denoted $P_{r1}$.
To compensate for the phase acquired by qubit~2, we then apply a rotation $R_Z(\theta) = \exp\left(-i\frac{\theta}{2}Z\right)$ on the second qubit, with $\theta := -\tpulse\DEZ$,
thus obtaining the following two-reference  gate:
 \begin{equation}
    P^\prime = (I \otimes R_Z(\theta))P_{r1}. \label{eq:Pr2}
\end{equation}

In the absence of noise ($\delta J = \wdeltak{1} = \wdeltak{2} = 0$), for an appropriate duration $\tpulse$ (see Section~\ref{subsec:simulation-parameters}) and assuming a sufficiently large $\DEZ / J_0$ ratio, this gate implements the following operator~\cite{meunier_efficient_2011}: 
\begin{equation}
    P^\prime=\left[\begin{array}{cccc}1 & 0 & 0 & 0 \\ 0 & e^{i\phi_{1}} & 0 & 0 \\ 0 & 0 & e^{i\phi_{2}} & 0 \\ 0 & 0 & 0 & 1 \end{array}\right],
\end{equation}
with  $\phi_{1}+\phi_{2}= \pi$.
Since the higher energy state experiences a smaller phase shift, it induces an asymmetric phase difference
 $\epsilon = \phi_1 - \phi_2$.
To implement the gate $P$ defined in~(\ref{eq:P_gate_def}), we must cancel this asymmetry, which can be done using either of the two approaches described below.

\subsubsection{Symmetry-corrected P gate}

For any $\DEZ$ value, the asymmetric phase shift difference $\epsilon$ can be determined  numerically, and then corrected by applying an appropriate rotation about the $Z$-axis  on each qubit. We get
\begin{equation}
P_{\text{sym-corr}} = \big( R_{Z_1}(\epsilon) \otimes  R_{Z_2}(-\epsilon)\big) P^\prime, \label{eq:P-sym-corrected}
\end{equation}
corresponding to the $P$ gate in~(\ref{eq:P_gate_def}), with subscript indicating the implementation method.

\subsubsection{$\pi$-pulse P gate}
\begin{figure}[h]
    \centering
\begin{minipage}{0.15\linewidth}\,\hfill\,\end{minipage}%
\begin{minipage}{0.8\linewidth}
\;\;\hfill\;\textcolor{Blue}{$\sqrt{P^\prime}$}\hfill\,\hfill\;\;\textcolor{Mahogany}{$X_1\otimes X_2$}\hfill\,\hfill\;\textcolor{Blue}{$\sqrt{P^\prime}$}\hfill\,\hfill\;\;\;\textcolor{Mahogany}{$X_1\otimes X_2$}\hfill\,
\end{minipage}\hfill\,\\
\begin{minipage}{0.15\linewidth}
\,\\[8mm]
\,\hfill\textcolor{Blue}{$J(t)/2$}\;\\
\,\hfill\textcolor{Mahogany}{$B_0(t)$\,\;}
\end{minipage}%
\begin{minipage}{0.8\linewidth}\includegraphics[width=\linewidth]{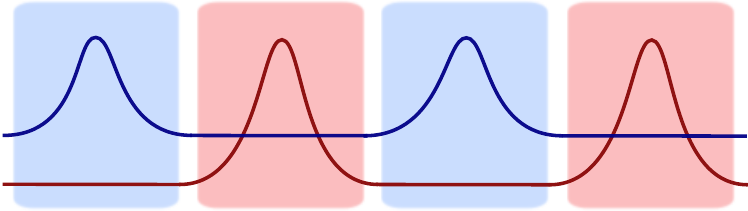}
\end{minipage}\hfill\,
    \caption{$J(t)$ and $B_0(t)$ pulses representation during a $\pi$-pulse $P$ gate. $J(t)/2$ indicates a $J(t)$ pulse with the same duration as for the $P^\prime$ gate, but with amplitude $J_0/2$. }
    \label{fig:P_pi_puse}
\end{figure}

Alternatively, a spin refocusing sequence can be used to implement the two-qubit gate $P$ (see also~\cite{tanttu2023assessment, barthel2023robust} for a similar use of spin refocusing pulses).
This approach has the advantage of being independent of the value of $\epsilon$ and therefore can be faithfully implemented even if $\epsilon$ is not known and/or slowly drifts over time. Moreover, using a spin refocusing sequence  also reduces the overall  phase shift acquired by each qubit during the gate operation.
The gate consists of a sequence of four pulses, as shown in Fig.~\ref{fig:P_pi_puse}, implementing the $P$ gate from~(\ref{eq:P_gate_def}):
\begin{equation}
P_{\pi\text{-pulse}} = (X_1 \otimes X_2) \sqrt{P^\prime} (X_1 \otimes X_2) \sqrt{P^\prime},
\end{equation}
where subscript $\pi$-pulse is used to indicate the implementation method.
To implement the  $\sqrt{P^\prime}$ gate, we use a $J(t)$ pulse with the same $\tpulse$ duration as for the $P^\prime$ gate, but with half the amplitude $J_0$ (in order to preserve an adiabatic process).
The $X$ gates are implemented as explained in Section~\ref{subsec:gates-with-driving-field}.

Finally, we note that for noisy gates, the Hamiltonian $H_{r1}$ is integrated numerically, and the rotation or $\pi$-pulse ($X$) gates used to implement the $P_{\text{sym-corr}}$ or  $P_{\pi\text{-pulse}}$ gates are also assumed to be noisy, following the single-qubit gate noise model  described earlier. The noise resilience of sym-correct and $\pi$-pulse gates is discussed in  Section~\ref{subsec:gates-fidelity} and their impact on the logical qubit performance is presented in Appendix~\ref{ap:complement-logical-qubit-perf}.

\subsection{Non-Markovian Noise Process}
\label{subsec:non-markovian-noise}

Silicon-based spin qubits are predominantly affected by non-Markovian stochastic processes governing $\wdelta$ and $\delta \VE$ deviations~\cite{chan_assessment_2018, struck_low-frequency_2020}. A  non-Markovian stochastic process exhibits autocorrelations that can be described in the frequency domain by the power spectral density (PSD) function. Precisely, for an ergodic, wide-sense stationary, real-valued, continuous-time stochastic process $x(t)$, the PSD function $S_{xx}(f)$ is given by
\begin{equation}
    S_{xx}(f) = \int_{-\infty}^{\infty} R_{xx}(\tau) e^{-i2\pi f \tau}\, d\tau,
\end{equation}
where $R_{xx}(\tau) =  \lim_{T\rightarrow \infty} \frac{1}{T}\int_{0}^{T} x(t)x(t+\tau)\,dt$
is the autocorrelation function. 

In the following, we  denote a PSD function as $S(f)$, with the corresponding stochastic process being clear from the context.  

We now discuss the laws governing the deviations $\wdelta$ and $\delta J$, affecting the one- and two-qubit gates presented above. The Larmor frequency fluctuates due to both charge noise and nuclear spin bath fluctuations,  which can be modeled as interactions with fluctuating two-level systems (TLSs). Each TLS individually generates Brownian noise, characterized by a PSD function $S(f)$ proportional to $1/f^2$. 
However, the collective effect of multiple TLSs can leads to a noise spectrum  $S(f) = S_0/f^\alpha$, with $\alpha \in [0,2]$, depending on the proximity and coupling strength of the TLSs in the device \cite{chan_assessment_2018, struck_low-frequency_2020}. Parameter $S_0$ is referred to as noise intensity and $\alpha$ as spectral exponent.

To characterize the noise in the coupling parameter $J$, we  consider its dependence on the  electrical potential $\VE$ of the exchange gate, which  determines both the height and width of the potential barrier. This leads to an exponential dependence of $J$ on $\VE$~\cite{qiao_coherent_2020}, which can be expressed as
\begin{equation} \label{eq:J-VE-exp-fit}
J(t) = a\exp\big(b\VE(t)\big),
\end{equation}
where $a$ and $b$ are fitting parameters. We use data from~\cite{xue_quantum_2022} to fit this exponential dependence;  details are provided in Appendix~\ref{ap:mV_to_J}. Since charge noise also causes deviations $\delta \VE$ in the exchange gate potential, using \eqref{eq:J-VE-exp-fit}, we get an additional exchange contribution in \eqref{eq:2q_H} with strength
\begin{equation}\label{eq:delta-J(t)}
    \delta J(t) 
    = J(t) \, (\exp(b \delta \VE(t))-1).
\end{equation}

In the following, we model both $\wdelta$ and $\delta \VE$ deviations as pink noise, that is, with PSD function proportional to $1/f$~\cite{chan_assessment_2018, struck_low-frequency_2020}, or in our terminology $\alpha = 1$.
These deviations are thus characterized by their intensity parameters.
The $\wdelta$ noise affects both single- and two-qubit gates, while $\delta \VE$ is converted into $\delta J$ noise, which specifically impacts two-qubit gates.

The next section details the approach used to emulate the effects of $\wdelta$ and $\delta \VE$ noise during the surface code error correction process.

\section{Spin-Qubit Surface Code Emulation}\label{sec:qec_emulation}

The surface code is widely regarded as one of the most promising QEC candidates for scaling up quantum devices, due to its ability to tolerate high  physical error rates and efficient implementation on architectures with nearest-neighbor connectivity only~\cite{kitaev_fault-tolerant_1997, dennis_topological_2001, raussendorf_fault-tolerant_2006, bombin_quantum_2007}. In particular, it aligns with our hardware model from Section~\ref{sec:hardware_noise_model}. Here, we focus on the emulation of a \mbox{distance-3} rotated surface code~\cite{fowler_surface_2012, tomita_low-distance_2014}, comprising 17 physical qubits (9 data  and 8 ancilla qubits), which is illustrated in Fig.~\ref{fig:rotated_surface17}. Additionally, we also consider the XZZX variant of the rotated surface code~\cite{bonilla_ataides_xzzx_2021}, because it is known to be more robust against biased noise. 

To emulate the surface code while assuming Si-qubit native gates and the noise models from Section~\ref{sec:one_qubit_gate} and Section~\ref{sec:two_qubit_gate}, we proceed as follows: (i) We convert the syndrome measurement circuit in terms of native gates, and (ii) we implement discrete-time stochastic  noise processes to represent $\wdelta(t)$ and $\delta \VE(t)$ progression over time, and use their random realizations to define  noisy gates.  These steps  together with the detailed simulation setup are presented in the next sections. 

\subsection{Syndrome Measurement Circuit}

\begin{figure}[!t]  
    \centering
    \includegraphics[width=\linewidth]{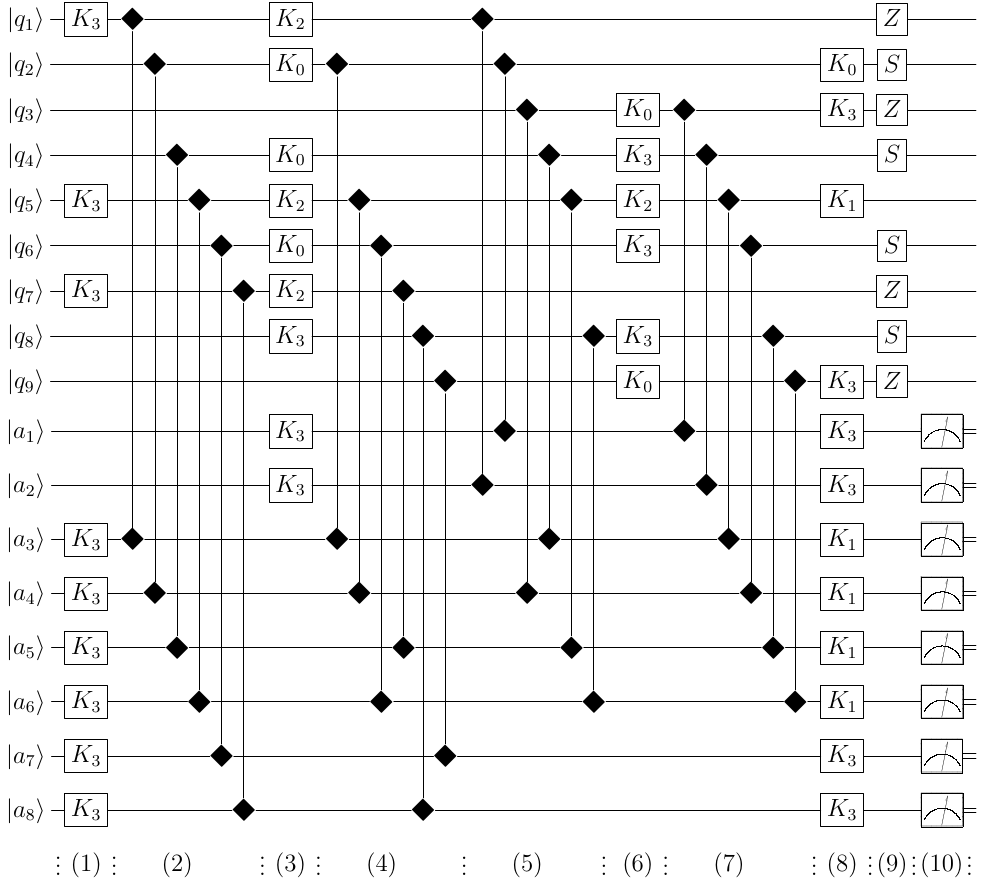}
    \caption{Native gates-based syndrome extraction circuit   for the rotated surface-17 code. The two-qubit $P$ gate is depicted by black filled  diamonds on the corresponding qubits, connected by a vertical line. 
    To ensure compatibility with the electronic limitations of Si-based quantum hardware, the circuit is divided into $10$ \emph{manipulation steps}, where gates of the same type are grouped together. We use the post-measurement state of the ancilla qubits for the subsequent syndrome extraction  (since we do not consider measurement errors, see Section~\ref{subsec:simulation-setup})), thus the initialization of the ancillas is not counted as a manipulation step.  }
    \label{fig:rotated_qcircuit}
\end{figure}

For any stabilizer code, the syndrome measurement can be implemented by using single-qubit initialization and measurement operations, Hadamard ($H$) gates, and controlled-$X$ ($CX$), or alternatively controlled-$Z$ ($CZ$) gates. For the distance-3 surface code, we use the improved stabilizer measurement circuit from~\cite[Section~III.B]{tomita_low-distance_2014}, which is converted into Si-qubit native gates, by using the following conversion rules:
\begin{align}
    H &= K_1 Z = Z K_1^\dagger \label{eq:H=K1Z}\\
    CZ &= P(S^\dagger \otimes S^\dagger) = (S^\dagger \otimes S^\dagger)P
\end{align}
This generates a number of additional $S^\dagger$ and $Z$ gates, which can be ``pushed'' to the end of the syndrome measurement circuit, by using~(\ref{eq:H=K1Z}), the fact that $S, S^\dagger$, and $Z$ gates commute with $P$, and \mbox{$SK_n = K_{n+1\, (\text{mod} 4)}S$}. In the end, we get the syndrome measurement circuit shown in Fig.~\ref{fig:rotated_qcircuit}, comprising a number of $68$ gates, divided into $9$ time steps (plus two additional time steps for  initialization and  measurement). To meet the requirements of controllability and parallelization, each time step consists of gates of the same type. This enables the parallel application of single-qubit gates of the same type (\ie, gates with or without driving field), or two-qubit gates that do not share any common qubits. Note that some qubits may remain idle during a given time step. In our simulations we implement the full syndrome measurement circuit, including step (9), although the latter does not need to be implemented in a real system, as the propagation of $S$ and $Z$ gates through the circuit can be tracked classically.

A similar procedure is applied to the XZZX variant of the rotated surface code, resulting into a syndrome measurement circuit comprising the same number of time steps, but $72$ gates. 

\begin{figure}[!t]
    \centering
    \includegraphics[width=\linewidth]{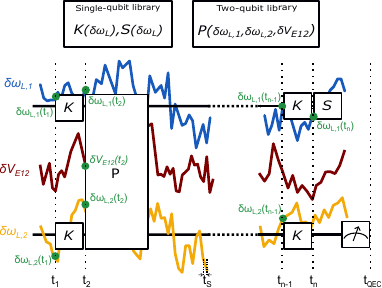}
    \caption{
    Construction of a noisy quantum circuit using the generated $\wdeltak{k}(t)$ and $\delta \VEkk{k_1 k_2}(t)$ time trace sequences. For each gate in the ideal circuit, we choose its corresponding noisy version from the precomputed library, according to $\wdeltak{k}(t)$ and/or $\delta \VEkk{k_1 k_2}(t)$ values at the start time $t$ of the gate. At each manipulation step, qubits are either idle or acted upon by gates of the same type (see Fig.~\ref{fig:rotated_qcircuit}), with no additional idle time inserted between consecutive manipulation steps (at step 9 in Fig.~\ref{fig:rotated_qcircuit}, qubits acted on by an $S$ gate are idle the second half of the step duration, which is determined by the duration of the $Z$ gate). This approach allows auto-correlated noise to be incorporated into the circuit, capturing the specific time-dependent variations.}
    \label{fig:simu_explanation}
\end{figure}

\subsection{Simulation Setup}
\label{subsec:simulation-setup}
We use the Qaptiva\textsuperscript{\sc tm} quantum emulator, to emulate  surface code error correction under the noise models described in Section~\ref{sec:native-gates}. We consider a Ramsey-like experiment, equivalent in this context to a quantum memory experiment, where an initial error-free logical state is protected through repeated syndrome measurements and error correction. To emulate the state evolution during noisy syndrome measurements, we proceed as follows:

\begin{enumerate}[leftmargin=12pt]

\item We precompute a large \emph{library} of noisy one- and two-qubit gates for a discrete range of values $\wdelta \in [-{\wdelta}_{\text{max}}, {\wdelta}_{\text{max}}]$ and $\delta \VE \in [-\delta {\VE}_{\text{max}}, \delta {\VE}_{\text{max}}]$. We make the assumption that  $\wdelta$ and $\delta \VE$ values remain constant during the gate.
For one-qubit gates, we set ${\wdelta}_{\text{max}}= 1$\,MHz and consider $10^5$ evenly spaced $\wdelta$ values. For two-qubit gates, we set $({\wdelta}_{\text{max}} = 0.8\,\text{MHz}, \delta {\VE}_{\text{max}} = 13\,\text{mV})$ or $({\wdelta}_{\text{max}} = 0.08\,\text{MHz}, \delta {\VE}_{\text{max}} = 6\,\text{mV})$, depending on the noise intensity, and consider $161$ evenly spaced $\wdelta$ values and $111$ evenly spaced $\delta \VE$ values, resulting in $\approx 2.9 \times 10^6$  $(\wdeltak{1}, \wdeltak{2}, \delta \VE)$ triplets. 

 To explicitly indicate the dependence of  a noisy gate on the $\wdelta$ and/or $\delta \VE$ values, we shall write $K(\wdelta)$, $S(\wdelta)$, $Z(\wdelta)$ for one-qubit gates, and $P(\wdeltak{1}, \wdeltak{2}, \delta \VEkk{12})$ for the two-qubit gate. Note that we indicate the dependency of the $P$ gate on $\delta\VE$ rather than $\delta J$, as the latter does not remain constant during the gate due to the exponential dependence of $J$ on $\VE$; see \eqref{eq:delta-J(t)}.

\item We consider a \emph{machine time} $t_{\text{m}}$, sufficient to accommodate a given (large) number of consecutive syndrome extractions. We also consider a sampling time $t_{\text{s}}$ (used to generate discrete noise sequences, see below), which is taken to be smaller than the duration of any one- or two-qubit gate (we actually take it to be a common divisor).

\item We generate discrete-time noise sequences, referred to as time trace sequences,  with length $t_{\text{m}}$ and sampling time $t_{\text{s}}$, as follows: one time trace sequence $\wdeltak{k}(t)$ is generated for each qubit $k$, and one time trace sequence $\delta \VEkk{k_1 k_2}(t)$ is generated for each pair of qubits $(k_1, k_2)$ acted on by a two-qubit gate (that is, for each pair consisting of an ancilla and a neighbor data qubit). We generate $1/f$ noise by a Fourier filtering method (see~\cite[Appendix E.1]{Tuorila2019}) and multiply the signal by a prefactor $\sqrt{S_0}$ to tune the intensity of the noise.

\item We execute the quantum circuit corresponding to the given number of consecutive syndrome extractions, after replacing each gate by the corresponding noisy gate from our precomputed  noisy gates library, \eg, $K(\wdeltak{k}(t))$, $S(\wdeltak{k}(t))$,  $P(\wdeltak{k_1}(t), \wdeltak{k_2}(t), \delta \VEkk{k_1 k_2}(t))$, see Fig.~\ref{fig:simu_explanation}.
Qubits that are not acted upon by any gate at a given manipulation step  undergo idle noise. 

\item We perform independent decoding of $X$ and $Z$ errors on windows of three consecutive syndromes with two consecutive windows overlapping by one syndrome extraction (see~\cite[Section~V.B]{tomita_low-distance_2014}). We use the look-up table based implementation of the minimum-weight perfect-matching decoder from~\cite{tomita_low-distance_2014}, and for each decoding window we track the corresponding error correction in the Pauli frame (to avoid unnecessary noisy quantum operations). After each decoding window, we compute the fidelity between the corrected state and the initial logical state, giving us one fidelity value for each $t=2\tQ, 4\tQ,\dots, 2n\tQ$, where $\tQ$ is the duration of one syndrome extraction and $n=\lfloor t_{\text{m}}/2\tQ\rfloor$.

\item We repeat steps 3-5 five hundred times, allowing us to determine the average fidelity between the corrected state and the initial logical state, as a function of time (that is, averaging, for each $t=2\tQ, 4\tQ,\dots, 2n\tQ$,  over fidelity values obtained at step 5).

\end{enumerate}

We note that state preparation (ancilla initialization) and measurement errors are intentionally excluded in order to accurately isolate the impact of non-Markovian noise on the fidelity of the logical qubit.
Moreover, since measurement errors are not considered in our simulations, we do not need to reinitialize the ancilla qubits at each syndrome extraction. Instead, we use their post-measurement state and track the corresponding changes in their Pauli frame.

\subsection{Spatial Noise Correlation}
\label{subsec:spatial-noise-correlation}

The non-Markovian property of noise discussed in Section~~\ref{subsec:non-markovian-noise} accounts for specific auto-correlations (in time) of the noise process. However, deviations of the Larmor frequency and exchange gate potentials may also be subject to spatial correlations, due to the shared environment of qubits~\cite{yoneda_noise-correlation_2023, rojas-arias_spatial_2023}. While a detailed investigation of refined correlation patterns lies beyond the scope of this work, we nonetheless consider two boundary cases: \emph{(i)}  spatially uncorrelated noise and \emph{(ii)} fully spatially correlated noise. Our simulation results will show that considering only these two boundary cases is sufficient, given the relatively small size of the QEC code. 

\emph{(i)} Spatially uncorrelated noise corresponds to the simulation setup described in Section~\ref{subsec:simulation-setup}: noise sequences $\wdeltak{k}(t)$ and  $\delta \VEkk{k_1 k_2}(t)$  are generated independently for each qubit $k$, or pair of qubits $(k_1, k_2)$ acted on by a two-qubit gate. 

\emph{(ii)} For fully spatially correlated noise, we only generate one $\wdelta(t)$ sequence and one   $\delta \VE(t)$ sequence, each with a specified intensity. Then we set $\wdeltak{k}(t) = \wdelta(t)$ for any qubit $k$, and $\delta \VEkk{k_1 k_2}(t) = \delta \VE(t)$ for any pair of qubits $(k_1, k_2)$ acted on by a two-qubit gate.

\begin{figure}[!b]
    \centering
    \includegraphics[width=0.7\linewidth]{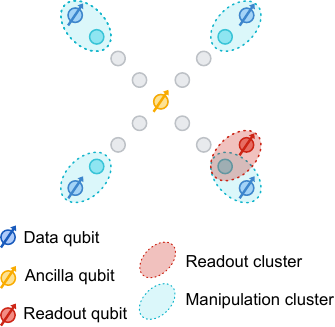}
    \caption{Example of a cluster-based architecture. Manipulation clusters (MCs) represent zones of high wiring density,  enabling  fine-tuning of QD Larmor frequencies, metal wiring for  ESR implementation, and exchange gates for  two-qubit operations. Readout clusters (RCs) incorporate  extra qubits for spin-to-charge conversion. MCs and RCs can be adjacent, as shown in the figure, or remotely located, with the architecture characterized by the distance between clusters.}
    \label{fig:cluster_architecture}
\end{figure}

\subsection{Sparse Architecture}\label{sec:sparse_architecture}

Thus far, we have considered the conventional qubit layout for implementing the surface code, which consists of a 2D grid of qubits with nearest-neighbor connectivity. However, such a dense architecture faces significant challenges due to wiring fan-out, substantial qubit cross-talk, or the need to accommodate additional qubits  for spin-to-charge conversion based readout. The realization of coherent shuttling~\cite{fujita2017coherent, flentje2017coherent, jadot2021distant, seidler2022conveyor, van2024coherent} enables spatially distributed architectures, where manipulation, initialization, and readout are performed in dedicated regions, facilitating more scalable and less complex wiring schemes.
Various architectures have recently been proposed in the literature, including the Spiderweb array~\cite{boter_spatial_2020} and SpinBus~\cite{kunne_spinbus_2024} architectures.
Aligning  with these approaches, we consider a generic sparse architecture consisting of manipulation clusters (MCs) and readout clusters (RCs) connected by 1D arrays of quantum dots (Fig.~\ref{fig:cluster_architecture}). In such an architecture, ancilla qubits must be shuttled between clusters, where they can be manipulated, interact with data qubits, or be read out. 
To simulate such an architecture, we introduce shuttling steps in the syndrome measurement circuit. During a shuttling step, ancilla qubits are shuttled between MCs and RCs, while data qubits remain idle. The duration of a shuttling step depends on the distance between the clusters (\ie, the length of the 1D array of QDs connecting them), and the time required to move a qubit between two adjacent QDs.  
As a first approximation, shuttling noise is modeled as idle noise (dephasing). 
However, we introduce a multiplicative scaling factor to account for the potential contribution in the effective noise intensity experienced by the shuttled qubit. An increase in noise may result from low-fidelity shuttling operations or from situations involving high spin qubit variability~\cite{ferdous2018interface, cifuentes2024bounds}, while a reduction can arise from motional narrowing~\cite{mortemousque2021enhanced, langrock2023blueprint, struck2024spin}. (see Section~\ref{subsec:noise-corr-sparse-archi-results}).

\section{Performance Metrics}

\setlength{\abovecaptionskip}{4pt}

In this section, we detail the time-based quantitative metrics used to assess the performance of  physical (Section~\ref{sec:physical_noise}) and logical  (Section~\ref{sec:QEC_metrics}) qubits.

\label{sec:performance-metrics}

\subsection{Physical Noise}
\label{sec:physical_noise}

Spin qubit systems are characterized by two  key timescales: the relaxation time  $T_1$ and the dephasing time $T_2^*$. In silicon-based spin qubits, $T_2^*$ is typically the limiting factor, as it is two to three orders of magnitude shorter than $T_1$~\cite{chan_assessment_2018}. Since our noise model assumes that idle qubits are subject  to dephasing only, $T_1$ is effectively infinite and we  consider solely  $T_2^*$. We determine the $T_2^*$ value using a simulated Ramsey experiment, where the $\ket{+}$ state evolves freely for a variable delay time $t$, undergoing dephasing due to  pink   $\wdelta$ noise. The  average fidelity  between the state at time $t$ and the initial  $\ket{+}$ state, denoted as $f(t)$, is then fitted by a Gaussian function~\cite{struck_low-frequency_2020},
\begin{equation}\label{eq:f}
    f(t) \approx \frac{1}{2}\left( 1 + \exp \left( - (t/T_2^*)^2 \right) \right),
\end{equation}
determining the $T_2^*$ value. 
In our model, the value of $T_2^*$ depends on the intensity $S_0$ of the $\delta \omega$ noise and the machine time $t_{\text{m}}$ (\ie, the length of the generated noise sequences), and can be fitted by:  
\begin{equation}
    T_2^*(t_{\text{m}}, S_0) \approx \left( 2\pi \sqrt{S_0 (\ln{t_{\text{m}}} - C)} \right)^{-1},\label{eq:T2_func}
\end{equation}
for a suitable choice of $C$,
as detailed in Appendix~\ref{ap:ramsey_exp}. This expression is consistent with theoretical models presented in~\cite{struck_low-frequency_2020, yoneda_quantum-dot_2018}.  We also note that the approximate Gaussian decay of the fidelity is characteristic of more general non-Markovian $1/f^\alpha$ noise, rather than being specific to pink, $1/f$ noise~\cite{struck_low-frequency_2020}.

Similarly, the characteristic time of coupled two-qubit system, $T_J^*$, is defined by a Ramsey-like experiment on a two-qubit system initialized in the $\ket{+}\ket{0}$ state. We set the exchange gate potential $\VE$  to a value corresponding to the amplitude $J_0$ of the coupling pulse $J$, and let the two-qubit system evolve freely for a variable delay of time $t$, while undergoing a potential deviation $\delta \VE(t)$. We determine $T_J^*$ by fitting the decay of the average fidelity between the two-qubit state at time $t$ and the initial  $\ket{+}\ket{0}$ state, denoted as $f_J(t)$, by a Gaussian function~\cite{dial_charge_2013},
\begin{equation}\label{eq:fJ}
    f_J(t) \approx \frac{1}{2}\left( 1 + \exp \left( - (t/T_J^*)^2 \right) \right).
\end{equation} In Appendix~\ref{ap:ramsey_exp} we show that $T_J^*$ also exhibits an inverse square-root dependence on $\ln(t_{\text{m}})$, similarly to one-qubit systems~\eqref{eq:T2_func}.

In the remainder of the paper, we shall characterize the strength of the physical noise using the characteristic time pair $(T_2^*, T_J^*)$, rather than the PSD intensity of $\wdelta$ and $\delta \VE$ noise.

\subsection{Logical Noise}\label{sec:QEC_metrics}

\begin{figure}[!t]
    \centering
    \includegraphics[width=\linewidth]{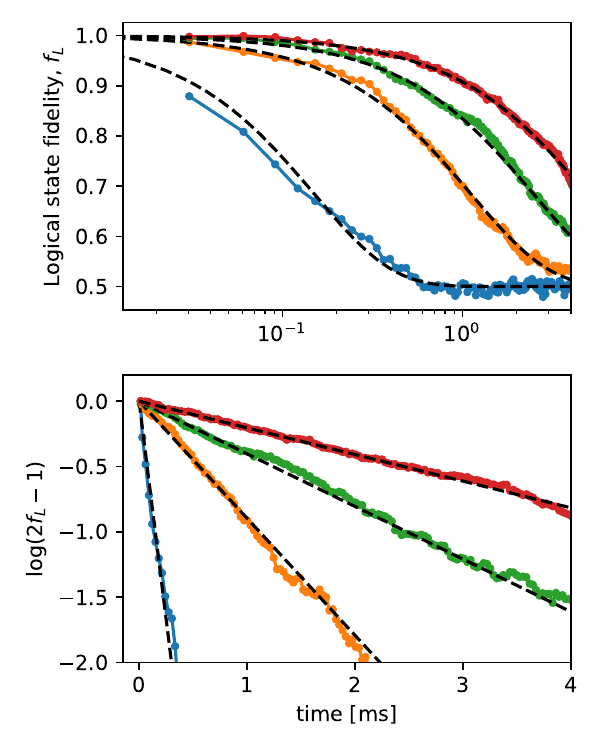}
    \caption{Average fidelity over time between the logical qubit state and the initial logical state $\ket{+}_L$ (top: linear scale, bottom: logarithmic scale). The different curves correspond to various $T_2^*$ values (the $T_J^*$ value is fixed). Black dashed curves represent the exponential fit of the average fidelity, giving the $\TL$ value, as explained in the text. Results have been obtain for \mbox{$S_0 \in \{10^{-5},\, 2.5\times10^{-6},\, 1.5\times10^{-6},\,  10^{-6} \} $\,MHz$^2$}. }
    \label{fig:logical_time_fit}
\end{figure}

 We define  characteristic times $T_{1,L}$ and $T_{2,L}^*$ for the logical qubit, providing performance metrics   directly comparable to the characteristic times of the physical qubit. To this end, we consider the quantum-memory experiment described in Section~\ref{subsec:simulation-setup}, where an initial error-free logical state, either $\ket{0}_L$ or $\ket{+}_L$, is maintained over time through repeated syndrome measurements and error correction. We then consider the decay of the average fidelity of the logical qubit as a function of time, denoted $f_{L}(t)$, analogous to the approach used in a Ramsey experiment. Importantly, and in contrast to the Gaussian decay happening for physical qubits, we observe that for logical qubits the average fidelity exhibits an exponential decay  (Fig.~\ref{fig:logical_time_fit}). 
 Such an exponential decay suggests that quantum error correction effectively Markovianizes the noise\,---\,that is, it transforms non-Markovian physical noise into Markovian logical noise\,---\, which can be seen as analogous to the way quantum error correction decoheres noise~\cite{greenbaum_modeling_2017, beale_quantum_2018}.

 Thus, we determine the characteristic time\,---\,$T_{1,L}$ if the initial logical state is $\ket{0}_L$, or $T_{2,L}^*$ if the initial logical state is $\ket{+}_L$\,---\,by performing an exponential fit of the average fidelity of the logical qubit as a function of time, \eg, 
 \begin{equation}\label{eq:fL}
     f_{L}(t) \approx \frac{1}{2}\left( 1 + \exp(-t/T_{2,L}^*) \right)
 \end{equation} for the logical qubit dephasing time.  In Appendix~\ref{ap:log_type_compare}, we show that $T_{2,L}^*$ values are lower than $T_{1,L}$, even though the syndrome measurement circuit slightly mitigates the dephasing bias of the physical noise model. Therefore, only the $T_{2,L}^*$ value will be reported in our numerical results (Section~\ref{sec:numerical-results}), as it limits the performance of the logical qubit. 
 
 Finally, we note that while $T_2^*$ and $\TL$ are useful metrics to  assess the performance of our QEC architecture, $\TL > T_2^*$ does not necessarily imply $f_L(t) > f(t)$, due to the difference between the Gaussian and the exponential fit in~\eqref{eq:f} and~\eqref{eq:fL}, respectively.

\section{Numerical Results}
\label{sec:numerical-results}

In this section, we first detail the simulation setup (Section~\ref{subsec:simulation-parameters}). We then present numerical results for single- and two-qubit gate fidelities (Section~\ref{subsec:gates-fidelity}), followed by an analysis of logical qubit performance (Section~\ref{sec:QEC_perfs}). Finally, we discuss the impact of spatial noise correlations and evaluate the performance of the surface code on a sparse architecture. (Section~\ref{subsec:noise-corr-sparse-archi-results}).

\subsection{Simulation Parameters}
\label{subsec:simulation-parameters}

\paragraph{Single-qubit gates:}
\label{subsec:simulation-param-1}
To maximize gates speed, we opt for the maximum feasible  values of $B_0$ and $\omega_{0}$ in \eqref{eq:gaussian-pulse} and \eqref{eq:cosine-pulse-omega}, respectively, that are compatible with the considered technology. In alignment with ESR qubit manipulation \cite{jacquinot2023rf}, we consider $B_0 \approx 2.1$\,MHz, resulting in $\tpulse = 1$\,$\mu$s to achieve a $\pi$-rotation about the $X$ or $Y$-axis. 
Additionally, we assume that the considered hardware can shift each qubit Larmor frequency up to $5$\,MHz, thus setting $\omega_{0} = 5$\,MHz, which leads to $\tpulse = 200$\,ns to achieve a $\pi$-rotation about the $Z$-axis.

\paragraph{Two-qubit gates:} $\DEZ$ and $J_0$ parameters are chosen as follows. 
(i) We set $\DEZ = 10$\,MHz, consistent with the previous assumption of $5$\,MHz controllability of qubit Larmor frequencies. 
(ii) We rely on numerical simulations to determine the maximum attainable $J_0$ amplitude that does not introduce non-adiabatic evolution effects (see Appendix~\ref{ap:pulse_shape_study}). 
For $\DEZ = 10$\,MHz,  our results indicate that $J_0 = 2$\,MHz is attainable, with a $J$-pulse duration  $\tpulse = 0.5\,\mu$s. For the symmetry-corrected $P$ gate, this leads to a gate time $t_{P_{\text{sym-corr}}} \approx  \tpulse = 0.5$\,$\mu$s, as the time taken by the $\epsilon$-rotation in~(\ref{eq:P-sym-corrected}) is negligible.  For the $\pi$-pulse $P$ gate, since a $\pi$-rotation about the $X$-axis takes $1$\,$\mu$s, we  obtain 
$t_{P_{\pi\text{-pulse}}} = 3\,\mu$s.

\paragraph{Measurement:} Although measurement errors are not considered in our simulations (to accurately capture the impact of non-Markovian noise on the logical qubit-performance, see Section~\ref{subsec:simulation-setup}), we still need to account for the  time taken by measurement operations, as data qubits are idle and undergo dephasing during this period. We consider a measurement time of $1$\,$\mu$s, consistent with recent results~\cite{takeda_rapid_2024}.

\paragraph{Noise generation:}

We generate discrete-time noise sequences of length $t_{\text{m}}\approx 1.68$\,s, which  covers a maximum number of $10^6$ consecutive syndrome measurement rounds. We consider a sampling frequency of $10$\,MHz ($t_{\text{s}} = 0.1\mu$\,s), consistent with the $1/f$ slope of the noise PSD and gates duration on the order of $1$\,$\mu$s. 

We note that the  values of both physical and logical  characteristic times reported in the sequel are computed using the $t_{\text{m}}$ value specified above (see also Appendix~\ref{ap:ramsey_exp} for the $T_2^*$ dependence on $t_{\text{m}}$).

\paragraph{Syndrome measurement time:} Considering the above simulation parameters (\emph{a}.-\emph{c}.), the syndrome measurement time is $\tQ\approx 5.2\,\mu$s for the symmetry-corrected $P$ gate, and $\tQ=15.2\,\mu$s for the $\pi$-pulse $P$ gate.

\subsection{Gates Fidelity}
\label{subsec:gates-fidelity}

Here, we present numerical results from simulations of the single- and two-qubit native gates discussed in Section~\ref{sec:native-gates}. 

\begin{figure}[!t]
    \centering
    \includegraphics[width=\linewidth]{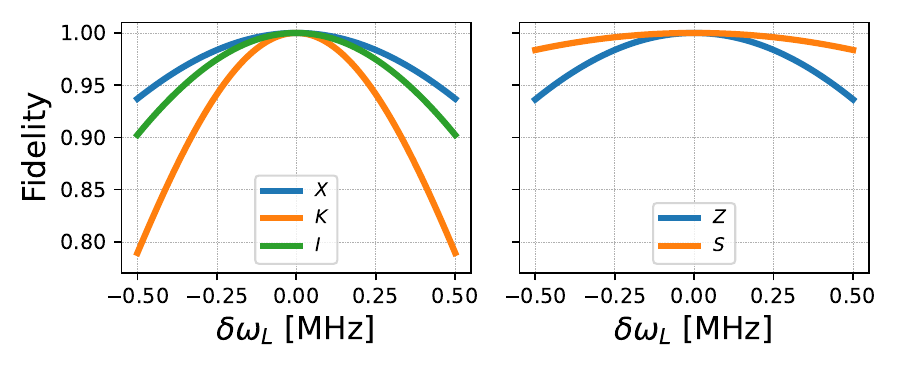}
    \caption{Impact of the Larmor frequency deviation $\wdelta$ on the gate fidelity for $X$, $K$ and idle (left) and for $Z$ and $S$ gate (right). Despite being 2 time longer, the X gate reveals better noise resilience (see main text) benefiting from spin-echo-like effect. For $Z$ and $S$ only the time of the gate impact the gate noise resilience.} 
    \label{fig:1qubit_X_K}
\end{figure}

Fig.~\ref{fig:1qubit_X_K} presents the fidelities of various single-qubit gates---namely $I$, $X$, $K$, $S$, and $Z$ gates ---as a function of the Larmor frequency deviation $\wdelta$. The idle gate $I$ corresponds to a qubit that remains idle for  the same duration as the $K$ gate ($0.5\,\mu$s). 
The left panel shows that despite having twice the duration of the $I$ and $K$ gates, the $X$ gate exhibits superior noise resilience, while the $K$ gate is noticeably noisier than the idle gate of the same duration. We note that for the $X$ and $K$ gates, which involve a driving field, two primary effects due to $\wdelta$ deviations can be identified: (i) the off-resonant driving, which leads to under-rotation of the qubit state, and (ii) the phase accumulation due to $\wdelta$. However, in the case of the $X$ gate the latter effect is partially mitigated by spin-echo-like behavior associated with rotations about the $Y$ or \mbox{$X$-axis}, resulting in a better noise resilience.
The right panel of Fig.~\ref{fig:1qubit_X_K} shows the fidelity of the $Z$ and $S$ gates. As expected, the $S$ gate demonstrates higher noise resilience, consistent with its reduced duration compared to the $Z$ gate.

\begin{figure}[!t]
    \centering
    \begin{tikzpicture}
        \node[anchor=south west, inner sep=0] (image) at (0,0) {
        \includegraphics[width=\linewidth]{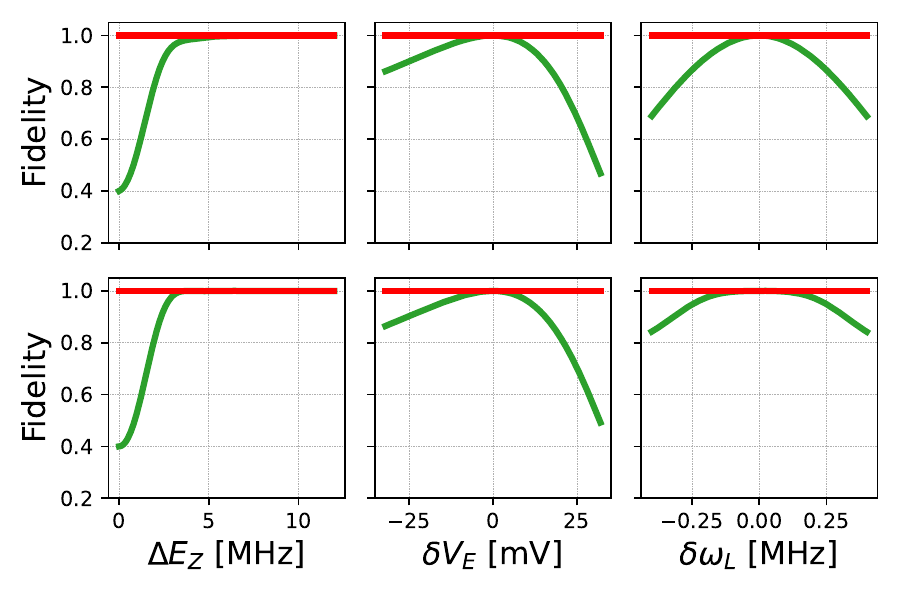}
        };
        \begin{scope}[x={(image.south east)},y={(image.north west)}]
            \node[anchor=north] at (0.02, 1.0)  { a)};
            \node[anchor=north] at (0.02, 0.575) { b)};
            ;
        \end{scope}
    \end{tikzpicture}
    \caption{Impact of the energy difference parameter $\DEZ$ (left), and of deviations $\delta\VE$ (middle) and $\wdelta$ (right) on the fidelity of the   symmetry-corrected $P$ gate  (a), and the $\pi$-pulse $P$ gate  (b). The Larmor frequency deviation  $\wdelta$ is applied to only one of the two qubits, while the second experiences no deviation from its reference.}
    \label{fig:P_gate}
\end{figure}

Fig.~\ref{fig:P_gate} presents the fidelity of the two $P$ gates from Section~\ref{sec:two_qubit_gate}, namely the symmetry-corrected $P$ gate, and the $\pi$-pulse $P$ gate, depending on  parameters  $\DEZ$ (left), $\delta \VE$ (middle), and $\wdelta$ (right), where the latter deviation is applied to only one of the two-qubits. The left panels corroborate the choice of gate parameters $J_0=2$\,MHz and $\DEZ = 10$\,MHz, as the gate fidelity closely approaches $1$ for $\DEZ \geq 5$\,MHz.  The middle panels reveal a higher impact of positive $\delta \VE$ deviations on the  fidelity of the two gates, explained by the exponential dependence of $J$ on $\VE$ potential. They also show that both the symmetry-corrected and  $\pi$-pulse $P$ gates achieve practically the same fidelity with respect to the $\delta\VE$ value. The two right panels highlight the advantage of the spin refocusing sequence applied for the $\pi$-pulse $P$ gate, particularly in mitigating the effect of $\wdelta$ deviations. Even though a single $\pi$-pulse ($X$) gate takes twice as long as the symmetry-corrected $P$ gate, the $\pi$-pulse $P$ gate exhibits a higher fidelity, due to the phase cancellation effect induced by the $\pi$-pulse. 

Note that this advantage becomes less important for higher $\wdelta$ values, as shown in Appendix~\ref{ap:pulse_shape_study}, Fig.~\ref{fig:pulse_2q}.

\subsection{QEC Performance}\label{sec:QEC_perfs}
Throughout this section, we consider the 17-qubit (distance-3) rotated surface code, implementing syndrome measurement using the $\pi$-pulse $P$ gate. Additional simulation results are provided in Appendix~\ref{ap:complement-logical-qubit-perf}, covering the symmetry-corrected $P$ gate and the XZZX variant of the rotated surface code. For completeness, we note that using the $\pi$-pulse $P$ gate yields better performance than its symmetry-corrected counterpart, except in the absence of $J$ noise, where both gates induce comparable performance. Additionally, the rotated surface code and its XZZX variant show comparable performance, since the syndrome measurement circuit does not preserve the noise bias.

\begin{figure}[!t]
    \centering
    \includegraphics[width=\linewidth]{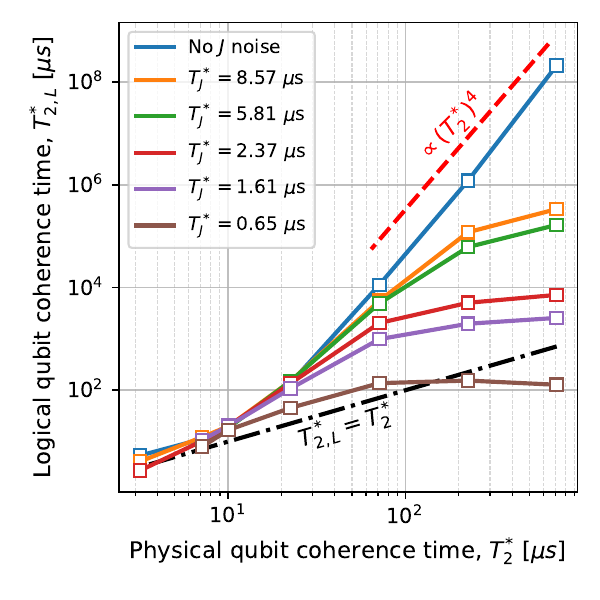}
    \caption{Logical qubit coherence time $T_{2,L}^*$ as a function of $T_2^*$, for different two-qubit coherence time values, $T_J^*$. In the absence of exchange energy ($J$) noise, the logical qubit coherence time scales as $T_{2,L}^* \propto(T_{2}^*)^{4}$, for $T_{2}^* \gtrsim 20\,\mu$s. We observe that low $T_J^*$ saturates the logical qubit coherence time, rendering improvements on $T_2^*$ ineffective.
    Error bars are computed based on a $95\%$ confidence interval of the average logical fidelity used to estimate $T_{2,L}^*$. The lower and upper bounds of this interval are then propagated to determine the corresponding minimum and maximum values of the coherence time, defining the error bars.}
    \label{fig:Jnoise_impact}
\end{figure}

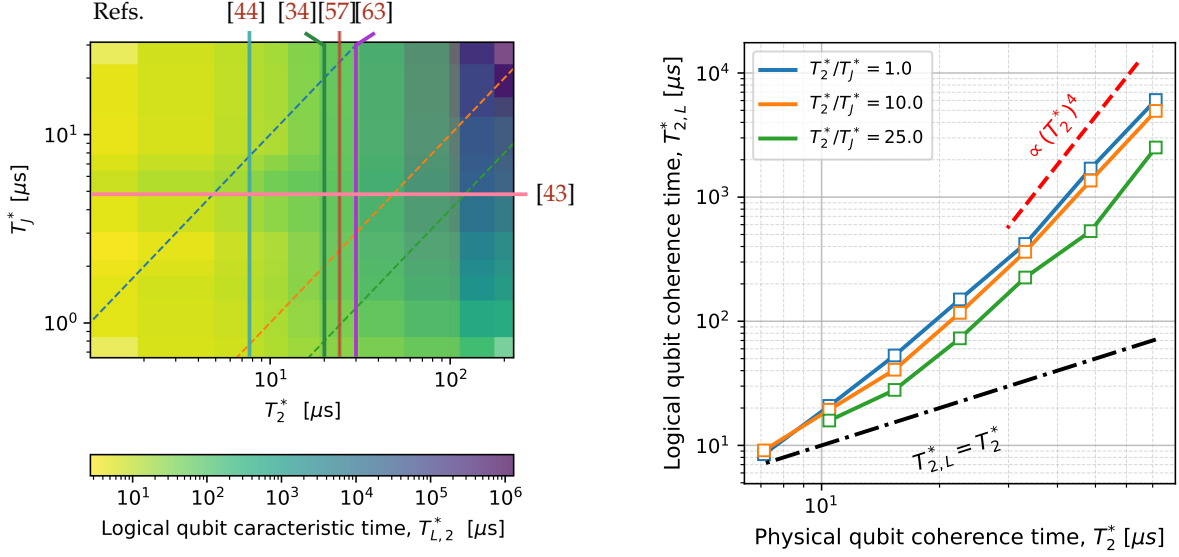
\begin{figure*}
    \centering
\,\hfill
\begin{tikzpicture}
        \node[anchor=south west, inner sep=0] (image) at (0,0) {
        \includegraphics[height=75mm]{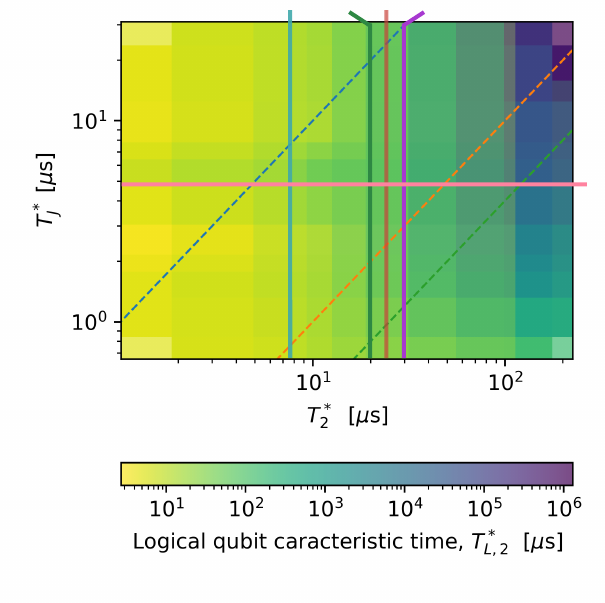}
        };
        \begin{scope}[x={(image.south east)},y={(image.north west)}]
            \node[anchor=north] at (0.25, 1.05) {Refs.};
            \node[anchor=north] at (0.475, 1.05) { \cite{rojas-arias_spatial_2023}};
            \node[anchor=north] at (0.565, 1.05) {\cite{struck_low-frequency_2020}};
            \node[anchor=north] at (0.635, 1.05) {\cite{yoneda_quantum-dot_2018}};
            \node[anchor=north] at (0.7, 1.05) { \cite{steinacker2024300}};
            \node[anchor=north] at (1.02, 0.727)  { \cite{yoneda_noise-correlation_2023}};
        \end{scope}
    \end{tikzpicture}\,\hfill
\includegraphics[height=75mm]{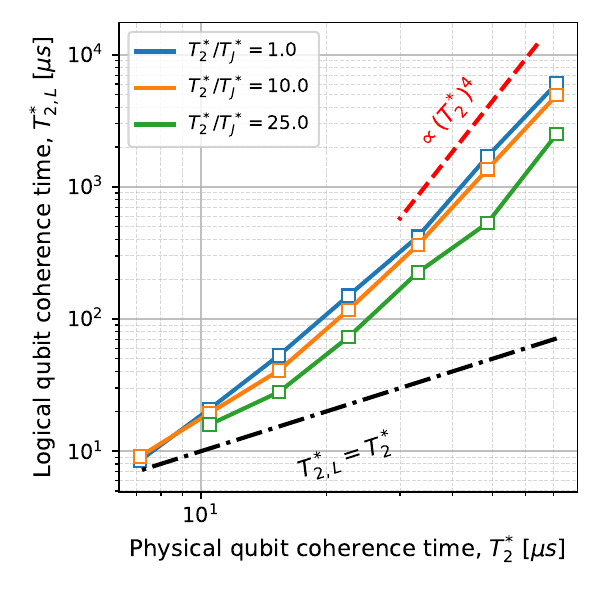}
\hfill\,
    \caption{(left) Heatmap visualization of $T_{2,L}^*$  as a function of $(T_2^*, T_J^*)$. Vertical and horizontal lines correspond to reference $T_2^*$ or $T_J^*$ values, as explained in the text. (right) $T_{2,L}^*$ as a function of $T_2^*$, assuming $T_J^*$ varies with $T_2^*$, such that the ratio $T_2^*/T_J^*$ remains constant. The three curves correspond to the three diagonal lines in the left panel.
    }
    \label{fig:joint_evolve}
\end{figure*}

\paragraph{Logical qubit coherence time.}

Fig.~\ref{fig:Jnoise_impact} shows the logical qubit coherence time $\TL$ as a function of the physical qubit  time  $T_{2}^*$, for various $T_J ^*$ values. 
In the absence of exchange energy noise, our numerical results reveal a quartic dependence of $\TL$ on $T_2^*$ (for sufficiently large  $T_2^*$), which arises from two factors: (1) the quadratic dependence on error rates (\ie, $1-$\,fidelity), corresponding to a code distance $d=3$, and (2) the transition from Gaussian decay of the fidelity at the physical level to exponential decay  at the logical level (Section~\ref{sec:QEC_metrics}; see also Fig.~\ref{fig:error_rate}). 

Precisely, let $p=1-f(\tQ)$ and $p_L=1-f_L(\tQ)$ denote the physical and logical error rates per QEC cycle, respectively.  
For $T_2^* \gg \tQ$, we have 
\begin{align}
p_L &\propto p^2, \quad \mbox{with }\ p_L \approx \frac{\tQ}{2\TL} \mbox{ and }   p \approx \frac{\tQ^2}{2(T_2^*)^2}, \label{eq:quartic-dependence-proof}
\end{align}
where the proportionality relation results from the code distance~\cite{fowler_surface_2012}, and the two  approximations are first order approximations arising  from the exponential and Gaussian decay of the fidelity, see~(\ref{eq:f}) and (\ref{eq:fL}), respectively.  Finally, (\ref{eq:quartic-dependence-proof}) implies:
\begin{equation} \label{eq:quartic-dependence}
\TL \propto \frac{(T_2^*)^4}{(\tQ)^3}. 
\end{equation} 
These results highlight a distinctive feature of non-Markovian noise: the quartic scaling seen with a distance-3 surface code, compared to the quadratic scaling under Markovian noise. 
Additionally, equation~(\ref{eq:quartic-dependence}) shows an inverse cubic dependence on  $\tQ$, highlighting  the substantial benefit of faster syndrome extraction.

\paragraph{Exchange energy noise impact.}
When exchange energy noise  is introduced, the error contribution from two-qubit gates becomes dominant at high $T_2^*$, due to the limitation on the $P$ gate fidelity, leading to a saturation of logical qubit performance as $T_2^*$ increases (Fig.~\ref{fig:Jnoise_impact}). We also note that in the absence of Larmor frequency noise, $\TL$ exhibits a quartic dependence on $T_J^*$ as shown in Appendix~\ref{ap:TJ_dependance}.

\paragraph{Heatmap and coupled noise variation.}

For a more comprehensive evaluation, Fig.~\ref{fig:joint_evolve} (left) provides a 2D heatmap visualization of $\TL$ as a function of $(T_2 ^*, T_J ^*)$. Each of the  three diagonal lines crossing the heatmap corresponds to a constant $T_2^*/T_J^*$ ratio.
Vertical lines indicate reference $T_2^*$ values reported in the literature.
It is important to note that these coherence times have been measured experimentally with various $t_m$ values and different effective PSD spectral exponents $\alpha$, which do not perfectly align with the $T_2^*$ calculation presented in Section~\ref{sec:physical_noise}.
The horizontal line corresponds to a $T_J^*$ value extracted from the PSD of $J$ fluctuations reported in~\cite{yoneda_noise-correlation_2023}, using  the methods described in Section~\ref{sec:physical_noise}.

The three curves in the right panel correspond to the three diagonal lines crossing the heatmap, with the $T_2 ^*/T_J ^*$ ratio value shown in the legend, and display the corresponding $\TL$ values as function of  $T_2^*$.
The ratio between the Larmor frequency and the $J$ coupling can vary depending on the hardware implementation, such as the use of ESR or micro-magnets. We note that the quartic dependence~(\ref{eq:quartic-dependence}) is still observed for various ratio values.

\subsection{Spatial Correlation and Sparse Architectures}
\label{subsec:noise-corr-sparse-archi-results}   
Here, we investigate the impact of the fully spatially correlated noise model from Section~\ref{subsec:spatial-noise-correlation}, and that of the sparse architecture considered in Section~\ref{sec:sparse_architecture}, on the performance of the 17-qubit rotated surface code, with syndrome measurement using the $\pi$-pulse $P$ gate. 

\begin{figure}
    \centering
    \includegraphics[width=\linewidth]{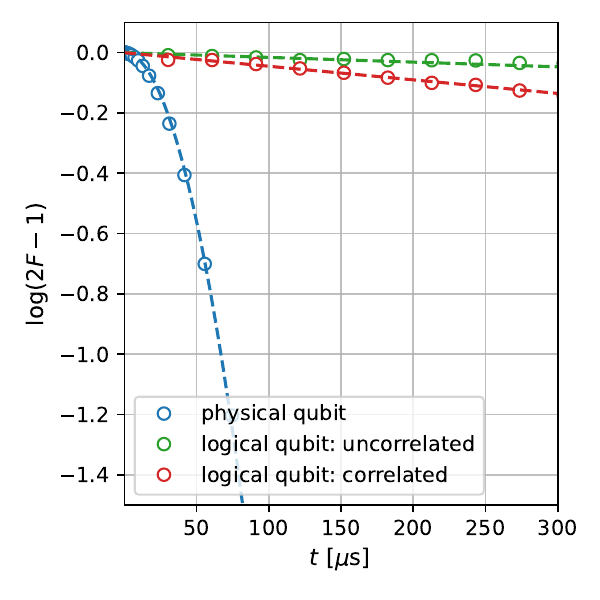}   
    \caption{
    Average fidelity of physical and logical qubits as a function of time,
    for  $T_2^* = 71.27\,\mu$s and without exchange energy noise.
    The curves plot $\ln(2F-1)$, where $F$ is the fidelity. For the logical qubit, both spatially uncorrelated noise and fully spatially correlated noise are considered. A Gaussian fit is applied to the physical qubit fidelity, while an exponential fit is used for the logical qubit. The insert zooms on small $t$ values, highlighting the difference between the Gaussian decay for the physical qubit and the exponential decay for the logical one.}
    \label{fig:error_rate}
\end{figure}

\paragraph{Spatial noise correlation.}

\begin{figure}
    \centering
    \includegraphics[width=\linewidth]{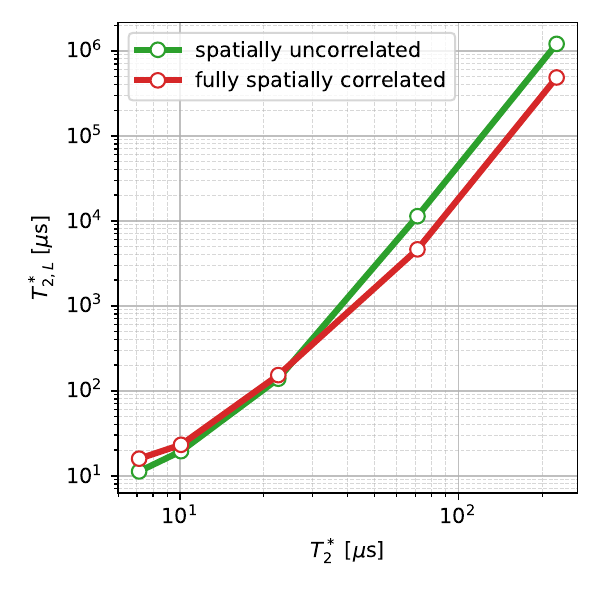}
    \caption{$T_{2,L}^*$ as a function of $T_2^*$ is presented for both spatially uncorrelated noise and fully spatially correlated noise scenarios, without exchange energy noise, for the rotated surface-17 code using a $\pi$-pulse $P$ gate. In the fully spatially correlated noise case, a quasi-quartic dependence is observed, demonstrating the low impact of fully spatially correlated noise in this configuration.}
    \label{fig:T2L_T2_correlation_graph}
\end{figure}

Fig.~\ref{fig:error_rate} shows the physical and logical qubit fidelity as a function of time, considering both spatially uncorrelated and  fully spatially correlated noise settings for the logical qubit. In the presence of spatially correlated noise, it can be observed that the fidelity decays exponentially with time, similar to the behavior observed under spatially uncorrelated noise. 
Thus, in both spatially uncorrelated and fully spatially correlated noise settings, we determine the coherence time of the logical qubit by using an exponential fitting or the fidelity decay.
Fig.~\ref{fig:T2L_T2_correlation_graph} shows the logical coherence time $\TL$ as a function of $T_2^*$, for both spatially uncorrelated and fully spatially correlated noise,in the absence of exchange energy noise. Fully spatially correlated noise has a limited impact, causing only a slight degradation in logical coherence time while maintaining a similar quartic dependence on $T_2^*$
  as observed in the spatially uncorrelated noise case for sufficiently high 
$T_2^*$.
This supports the fact that the syndrome measurement  mitigates the effect of spatial correlations, as each qubit may experience noise differently due to the applied gates and propagation effects (\ie, noise either affecting or propagating through gates of different types,  applied to different qubits at different time steps, see Fig.~\ref{fig:rotated_qcircuit}).

\paragraph{Sparse architecture.}
We consider the sparse architecture described in Section~\ref{sec:sparse_architecture}, where each ancilla qubit  shuttles through diagonal 1D arrays of QDs, interacting with corresponding data qubits in four distinct  manipulation clusters (Fig.~\ref{fig:cluster_architecture}). Each ancilla qubit has a dedicated readout cluster, which is assumed to be adjacent to the last manipulation cluster it visits. 

\begin{figure}
    \centering
    \begin{tikzpicture}
        \node[anchor=south west, inner sep=0] (image) at (0,0) {
        \includegraphics[width=\linewidth]{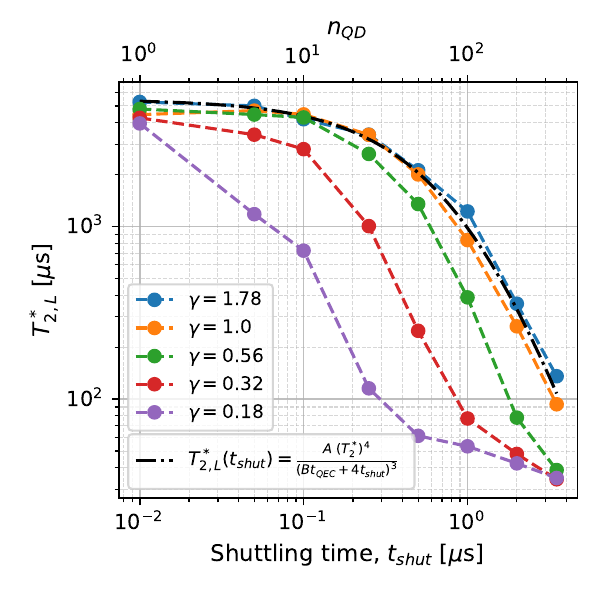}
        };
        \begin{scope}[x={(image.south east)},y={(image.north west)}]
            \node[anchor=north] at (0.340, 0.640)  { 
            \begin{tabular}{rl}
                    $T_2^*$ & $= 71.27\,\mu s$ \\
                    $T_J^*$ & $= 7.13\,\mu s$
            \end{tabular}
            };
        \end{scope}
    \end{tikzpicture}
    \caption{Logical qubit coherence time as a function of the shuttling time, $t_{\text{shut}}$ (bottom axis) and 1D array length, $\nQD$ (upped axis) for fixed physical coherence time $(T_2^*, T_J^*) = (71.27\mu s, 7.13\mu s)$.}
    \label{fig:shuttling_impact}
\end{figure}

The shuttling time between two manipulation clusters is given by $t_\text{shut} = \nQD\tau $, where $\nQD$ is the the length of the 1D array between the clusters, and $\tau$ is the shuttling time between two adjacent QDs. In our simulations, we take $\tau = 10$\,ns, consistent with a shuttling velocity of $v=10m/s$ \cite{langrock2023blueprint} and a QD spacing of the order of $100$\,nm \cite{elbaz2025transport}.
Fig.~\ref{fig:shuttling_impact} illustrates the impact of mapping the 17-qubit rotated surface code onto such a sparse architecture,  incorporating shuttling of ancilla qubits. We consider fixed coherence times $T_2^* = 71.27\,\mu\text{s}$ and $T_J^* = 7.13\,\mu\text{s}$, and vary the length of the 1D array between clusters (thus the shuttling time).

To account phenomenologically for the contribution of shuttling, we modifies the effective noise intensity experienced by the shuttled qubit. We also assume that this qubit frequency change occurs at a time scale shorter than the sampling time $t_s$. This leads to an effective reduction (or increase) of $T_2^*$ by a multiplicative factor $\gamma$, shown in the legend. 
For $\gamma=1$ (indicating the same noise intensity during shuttling), the logical coherence time can be fitted by:
\begin{equation}
\TL(t_\text{shut}) \approx A \frac{(T_2^*)^4}{(B\tQ + 4 t_\text{shut})^3},
\end{equation} 
with $A$ and $B$ fitting parameters. 
This follows from~(\ref{eq:quartic-dependence}), while replacing $\tQ$ by $B\tQ + 4 t_\text{shut}$, to account for the shuttling time during  syndrome extraction, and including a multiplicative factor $B$ in front of $\tQ$ (with $B \approx 0.34$), reflecting the fact that certain gates in the extraction circuit induce a partial spin-echo effect, effectively reducing the time during which the qubit is subject to decoherence.
As the value of $\gamma$  decreases, we observe that shuttling  degrades the logical error rate performance more rapidly.

Importantly, we observe that the QEC performance remains robust up to a shuttling time of approximately \mbox{$t_\text{shut} \approx 0.1\,\mu$s}.
Since both the shuttling time and the number of shuttling steps per syndrome extraction cycle are independent of the code distance, this robustness is expected to persist or even improve as the code distance increases.

\section{Conclusion}
\label{sec:conclusion}

We provided a comprehensive evaluation of the performance of the  distance-3 surface code in silicon-based spin qubits, under realistic non-Markovian noise conditions. Our results show that the coherence time of the logical qubit significantly exceeds that of the physical qubit, exhibiting quartic scaling in the presence of either Larmor frequency noise or exchange energy noise alone, as well as when both $T_2^*$ and $T_J^*$ evolves jointly.
Moreover, we analyzed the impact of spatially correlated noise and sparse qubit architectures, showing that the distance-3 surface code retains near-nominal error suppression capabilities (\ie, close to those observed  under idealized conditions) even under these constraints.

The methods presented here can be extended to higher-distance codes (\eg, $d=5$, and possibly $d=7$) by leveraging more efficient and sophisticated quantum simulators, such as those based on tensor networks. It would also be interesting to compare our results with those obtained from simulations using approximate Pauli noise models, such as those derived via the Pauli twirling approximation, and to derive the surface code threshold for these noise models. Our observation of logical noise Markovianization suggests that Pauli noise approximations could be sufficiently accurate despite the non-Markovian nature of the physical noise.

\section*{Acknowledgments}

OG and VS acknowledge funding from the Plan France 2030 through grant PRESQUILE (ANR-22-PETQ-0002) and the QuantERA grant EQUIP (ANR-22-QUA2-0005-01). TA and TM acknowledge funding from the European Union's Horizon 2020 research and innovation program, under grant agreement No951852 (QLSI project) 
BV acknowledges support from the hybdrid quantum initiative (HQI ANR-22-PNCQ-0002).

\appendix
\counterwithin{figure}{section}
\renewcommand\thefigure{\thesection.\arabic{figure}}

\clearpage
\section*{APPENDICES}

\section{Pulse Shape Study}\label{ap:pulse_shape_study}

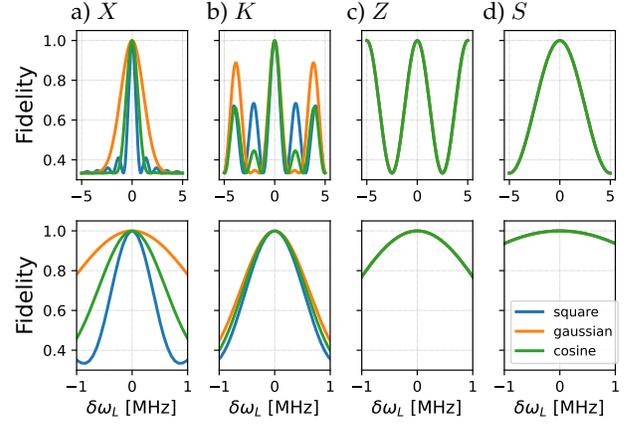
\begin{figure}[!t]
    \centering
    \begin{tikzpicture}
        \node[anchor=south west, inner sep=0] (image) at (0,0) {
        \includegraphics[width=\linewidth]{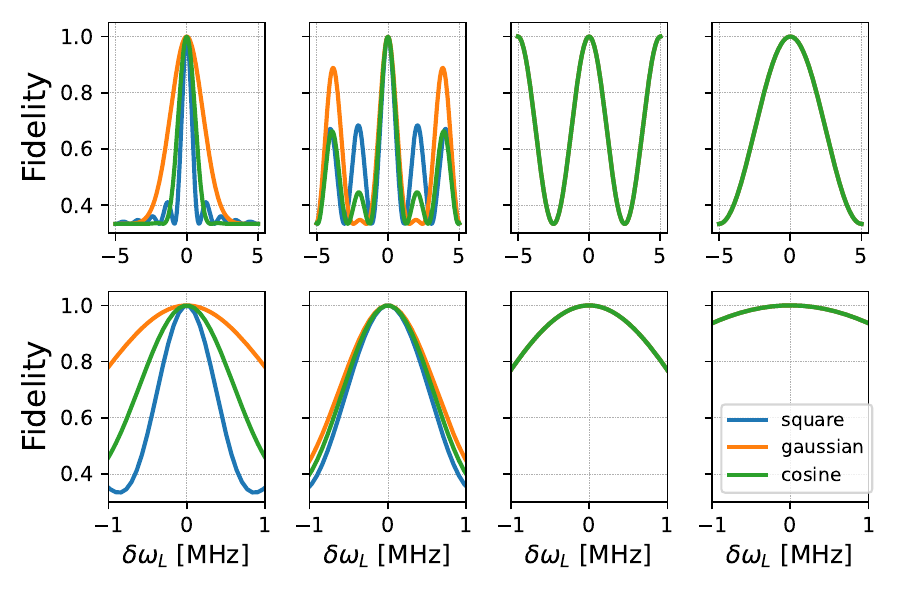}
        };
        \begin{scope}[x={(image.south east)},y={(image.north west)}]
            \node[anchor=north] at (0.15, 1.05)  {a) $X$};
            \node[anchor=north] at (0.364, 1.05) {b) $K$};
            \node[anchor=north] at (0.578, 1.05) {c) $Z$};
            \node[anchor=north] at (0.792, 1.05) {d) $S$};
        \end{scope}
    \end{tikzpicture}
    \caption{Impact of the Larmor frequency deviation $\wdelta$  on the  fidelity of single-qubit gates, for square, Gaussian and cosine pulses. a) $X$, b) $K$, c) $Z$, d) $S$.}
    \label{fig:pulse_1q}
\end{figure}

In this appendix, we describe the evaluation of gate performance across various single- and two-qubit gates implemented with different pulse shapes. We use numerical simulations to assess how gate fidelity depends on noise and to identify the most suitable pulse shapes.
See (\ref{eq:gaussian-pulse}) and (\ref{eq:cosine-pulse-omega}) for the Gaussian and cosine pulse equations respectively.
Fig.~\ref{fig:pulse_1q} illustrates the single-qubit fidelity as a function of Larmor frequency deviation $\wdelta$, with a fixed driving field amplitude $B_0$ for the $X$ and $K$ gates, and a fixed $\omega_0$ amplitude for the $Z$ and $S$ gates. 
As shown in the figure, using a Gaussian pulse yields superior performance for gates involving a driving field due to their broader frequency spectrum, enabling a wider driving frequency range. Consequently, Gaussian pulses are used in single-qubit gate with driving field and cosine pulse for single qubit gate without driving field enabling shorter pulse duration for  $Z$ and $S$ gates.

\begin{figure}[!t]
    \centering
     \begin{tikzpicture}
        \node[anchor=south west, inner sep=0] (image) at (0,0) {
         \includegraphics[width=\linewidth]{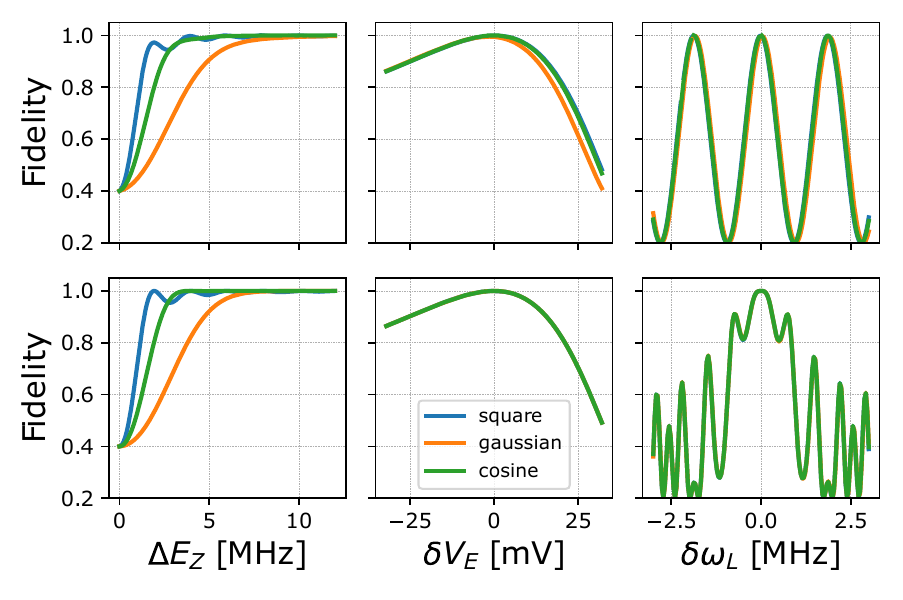}
        };
        \begin{scope}[x={(image.south east)},y={(image.north west)}]
            \node[anchor=north] at (0.02, 1.0)  { a)};
            \node[anchor=north] at (0.02, 0.575) { b)};
            ;
        \end{scope}
    \end{tikzpicture}
    \caption{Impact of  parameters $\DEZ$ (left), $\VE$ (middle) and $\wdelta$ (right) on the  fidelity of the $\pi$-pulse $P$ gate (a), and  symmetry corrected $P$ gate (b), for square, Gaussian and cosine $J$-pulse shapes.}
    \label{fig:pulse_2q}
\end{figure}

For a two-qubit $P$ gate, an adiabatic transformation is necessary to prevent flip errors, requiring a smooth rise-and-fall pulse shape, and a sufficiently large energy difference between qubits $\DEZ$ compared to the pulse amplitude $J_0$. Fig.~\ref{fig:pulse_2q} shows the impact of parameters $\DEZ$, $\delta \VE$, and $\wdelta$ on the gate fidelity, for different pulse shapes, where we take \mbox{$J_0 = 2$\,MHz} to achieve sufficiently fast $P$ gate.
As illustrated in the left panels, the cosine pulse enables better performance with a fast convergence to high fidelity as a function of $\DEZ$,  achieving a perfect gate for  \mbox{$\DEZ > 8$\,MHz}. Consequently, the cosine pulse is selected for the $P$ gate. In contrast, square and Gaussian pulses fail to provide steady robust  performance across a wide range of $\DEZ$ values, due to oscillations in fidelity for the square pulse and a slow-converging fidelity response for the Gaussian pulse.

\section{Conversion of Exchange Gate Potential to $J$ Coupling}\label{ap:mV_to_J}

\begin{figure}[!t]
    \centering
    \includegraphics[width=\linewidth]{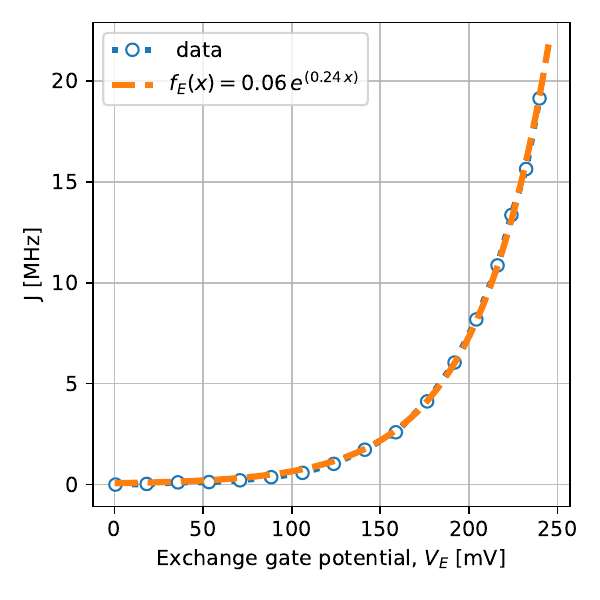}
    \caption{Dependence of the $J$ value [MHz] on the barrier pulse amplitude [mV]. Blue circle marks correspond to the experimental data from \cite{xue_quantum_2022}, which are fitted by the function $f_{\text{E}}$ shown in the legend.}
    \label{fig:mV_to_J_fit}
\end{figure}

The coupling parameter $J$ exhibits an exponential dependence on the exchange gate potential $\VE$~\cite{qiao_coherent_2020}. In Fig.~\ref{fig:mV_to_J_fit}, we use  data from~\cite{xue_quantum_2022} to fit this exponential dependence, and employ the fit function $f_{\text{E}}$ to
convert potential deviation $\delta \VE$ into  $\delta J$ values, depending on the time-varying $J$ pulse value (eq.~(\ref{eq:J-pulse-def}) in the main text):
\begin{multline}
\delta J(t) = f_{\text{E}}(\VE(t) + \delta\VE) - f_{\text{E}}(\VE(t)),\\ \mbox{ with } \VE(t) = f_{\text{E}}^{-1}(J(t)). 
\end{multline}

\section{Ramsey Experiments}\label{ap:ramsey_exp}
\begin{figure}[!t]
    \centering
    \includegraphics[width=\linewidth]{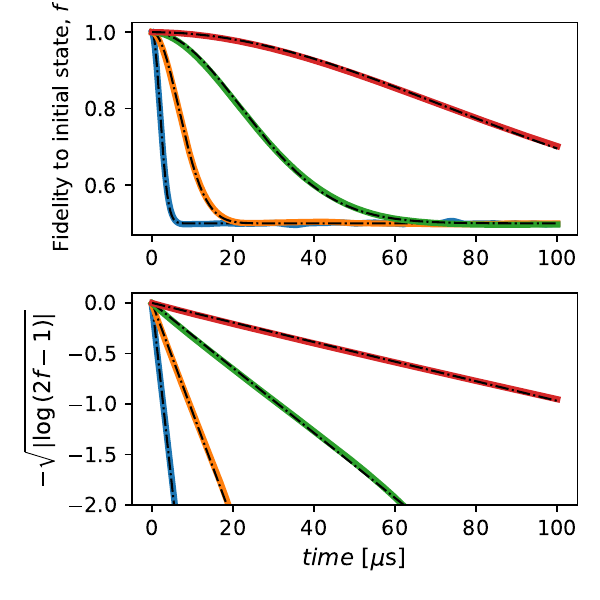}
    \caption{Average fidelity between qubit state at time $t$ and the initial $\ket{+}$ state, as a function of the free evolution time $t$. Solid curves correspond to different intensities of the $\wdelta$ noise for a fixed $t_{\text{m}} \approx 1.6\,$s with $S_0 \in \{2.5.10^{-4}, 2.5.10^{-5}, 2.5.10^{-6}, 2.5.10^{-7} \}$\,MHz$^2$.  Black dashed curves show the Gaussian fits used to determine $T_2^*$.}
    \label{fig:T2_fit}
\end{figure}

To determine the coherence time of the physical qubit, $T_2^*$, we generate $\wdelta$ noise sequences of length $t_{\text{m}}$, referred to as machine time, and use them to simulate Ramsey experiments. In these simulations, a qubit is prepared in the $\ket{+}$ state,  evolves freely for a time $t$  undergoing dephasing according to a generated  noise sequence, and then  its fidelity to the initial $\ket{+}$ state is recorded.
We then determine the average fidelity between the state at time $t$ and the initial  $\ket{+}$ state, denoted by $f(t)$, by averaging over different noise sequences. For $t_{\text{m}} \gg t$, the average fidelity $f(t)$ is closely approximated by~\cite{struck_low-frequency_2020, yoneda_quantum-dot_2018}:
\begin{equation}\label{eq:F-approx-yoneda}
    f(t) \approx \frac{1}{2}\left( 1 + \exp \left( - 4\pi^2 S_0 \ln(t_{\text{m}}/t) t^2 \right) \right).
\end{equation}
Since for $t_{\text{m}} \gg t > t_s$ (where $t_s$ is the sampling time), $\ln(t_{\text{m}}/t)$ varies only marginally with $t$, and one can further  approximate $f(t)$ by a Gaussian function:
\begin{equation}\label{eq:F-gauss-approx}
    f(t) \approx \frac{1}{2}\left( 1 + \exp \left( -  (t/T_2^*)^2 \right) \right).
\end{equation}
In practice, we determine the $T_2^*$ value  to minimize the root mean square error of the above approximation.  
Fig.~\ref{fig:T2_fit} shows the average fidelity $f(t)$ as a function of the free evolution time $t$, for a fixed machine time $t_{\text{m}} = 1.6\,$s, and different intensity values $S_0$ of the $\wdelta$ noise. It can be observed that $f(t)$ is closely matched by the Gaussian approximation~(\ref{eq:F-gauss-approx}) determining the $T_2^*$ value.

\begin{figure}[H]
    \centering
    \includegraphics[width=\linewidth]{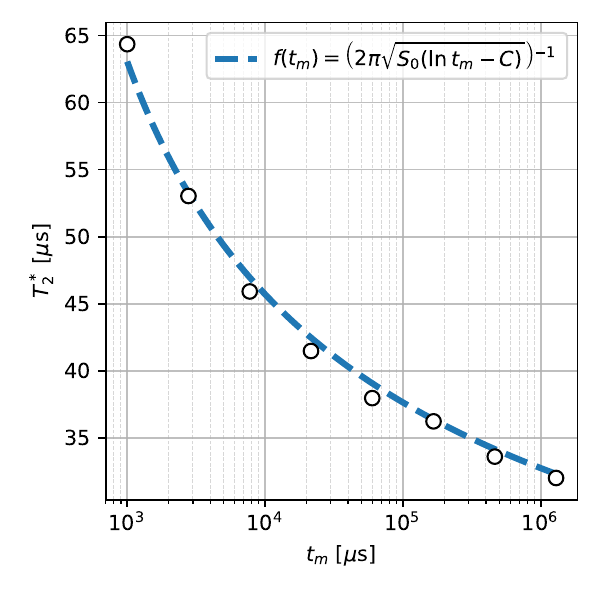}
    \caption{$T_2^*$ as a function of machine time, $t_{\text{m}}$, for fixed  intensity of the $\wdelta$ noise, $S_0=2.5.10^{-6}$\,MHz$^2$. Data points from numerical simulation are  fitted by $f(t_{\text{m}}) = \left( 2\pi \sqrt{S_0 (\ln{t_{\text{m}}} - C)} \right) ^{-1}$, for a suitable choice of $C \approx 4.36$, consistent with (\ref{eq:t2star_tm_approx}).}
    \label{fig:T2_tm}
\end{figure}

\clearpage
Equations~(\ref{eq:F-approx-yoneda}) and~(\ref{eq:F-gauss-approx}) also imply that $T_2^*$ can be closely approximated by
\begin{equation}
\label{eq:t2star_tm_approx}
T_2^* \approx \left( 2\pi \sqrt{S_0 \ln(t_{\text{m}}/t)} \right) ^{-1},
\end{equation}
for a suitable choice of $t$.
Hence, the $T_2^*$ value depends on the intensity $S_0$ of the $\wdelta$ noise,  as well as on the machine time $t_{\text{m}}$. 
The dependence on $t_{\text{m}}$ arises because a longer machine time increases the contribution of the low-frequency noise spectrum, resulting in noise realizations that are distributed over a wider range of values. This causes the qubit fidelity to decay faster, thus leading to a lower $T_2^*$ value. 
In Fig.~\ref{fig:T2_tm}, we plot the $T_2^*$ value obtained from numerical simulations, as a function of $t_{\text{m}}$. We observe that the dependence of $T_2^*$ on $t_{\text{m}}$ is consistent with the above approximation.

\begin{figure}[!t]
    \centering
    \includegraphics[width=\linewidth]{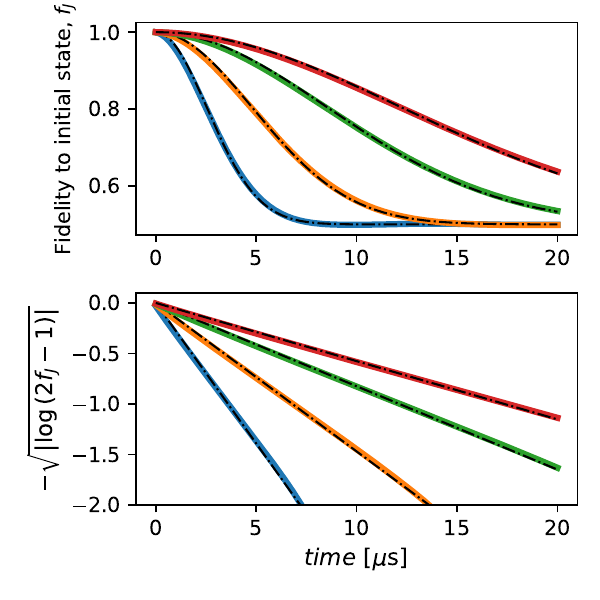}
    \caption{Fidelity to the initial state during a Ramsey experiment on a two-spin qubit system subject to $\delta J$ coupling fluctuations for various $\VE$ noise intensities, $S_0 \in \{0.25,   0.075,  0.025,  0.0125\}$\,MHz$^2$. The black dashed curves represent the Gaussian fit of the fidelity, yielding the corresponding $T_J^*$ values. }
    \label{fig:TJ_fit}
\end{figure}

\begin{figure}[!t]
    \centering
    \includegraphics[width=\linewidth]{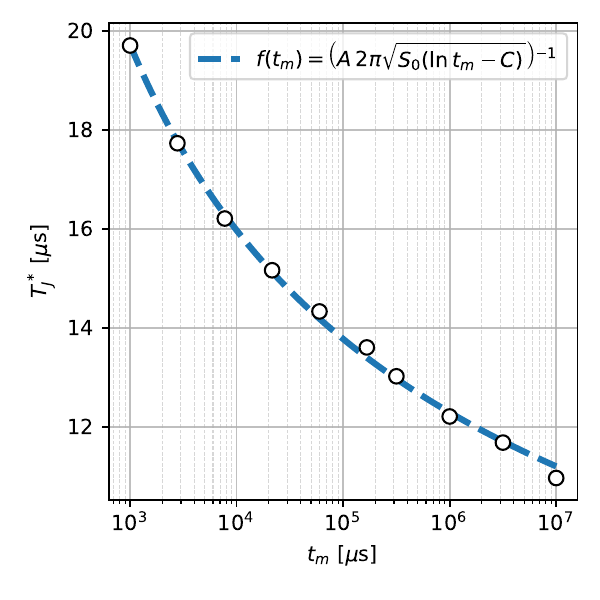}
    \caption{$T_J^*$ as a function of machine time, $t_{\text{m}}$, for fixed  intensity of the $\delta\VE$ noise, $S_0=0.025$\,MHz$^2$. Inverse square-root logarithmic dependence is observed on $t_{\text{m}}$, with data points from numerical simulation fitted by \mbox{$f(t_{\text{m}}) = \left( A\,2\pi \sqrt{S_0 (\ln{t_{\text{m}}} - C)} \right) ^{-1}$}, with \mbox{$(A,C) \approx (0.024,2.52)$.}}
    \label{fig:TJ_tm}
\end{figure}

 Similarly, Ramsey experiments were conducted on a two-spin qubit system, where the qubits undergo coupling fluctuations  $\delta J$ due to exchange gate potential noise $\delta\VE$ (see details in the main text). The characteristic time for such a system, $T_J^*$, is obtained through a Gaussian fit of the average fidelity decay, given by~(\ref{eq:fJ}).
Fig.~\ref{fig:TJ_fit} illustrates the results for Ramsey experiments with $t_{\text{m}} = 1.6$\,s and various $\delta \VE$ noise intensities. In Fig.~\ref{fig:TJ_tm}, we show that $T_J^*$ exhibits an inverse square root dependence on $\ln(t_{\text{m}})$, similar to one-qubit systems.

\section{Complementary QEC Evaluations}
\label{ap:complement-logical-qubit-perf}

\subsection{Logical Characteristic Times $T_{1,L}$ and $T_{2,L}^*$}\label{ap:log_type_compare}

\begin{figure}[!t]
    \centering
    \includegraphics[width=\linewidth]{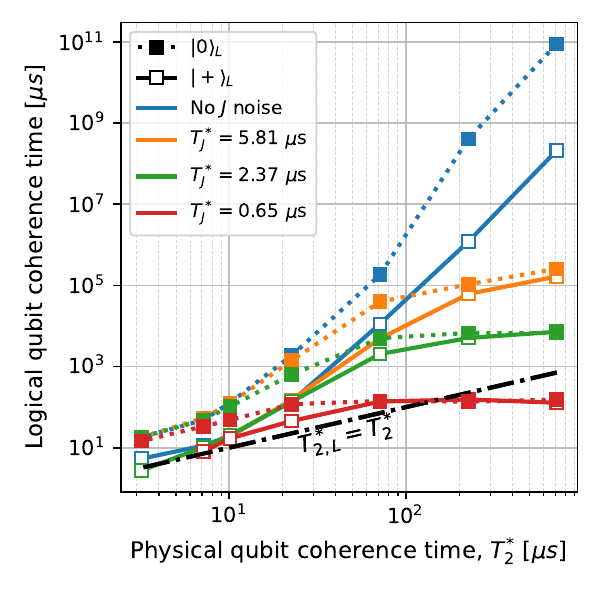}
    \caption{Logical qubit characteristic times $T_{1,L}$ and $T_{2,L}^*$, corresponding to logical qubits $\ket{0}_L$ and $\ket{+}_L$, respectively, as a function of $T_2^*$, for different $T_J^*$ values induced by exchange energy noise. Syndrome measurement use the $\pi$-pulse $P$ gate. For all  considered $(T_2^*,T_J^*)$ values, $\ket{+}_L$ exhibits lower characteristic time  compared to $\ket{0}_L$.}
    \label{fig:log_type_compare}
\end{figure}

Fig.~\ref{fig:log_type_compare} shows the characteristic times $T_{1,L}$ and $T_{2,L}^*$, corresponding to logical qubits $\ket{0}_L$ and $\ket{+}_L$, respectively, for various ($T_2^*$, $T_J^*$) values. It can be observed that  $\ket{+}_L$ exhibits a lower characteristic time  compared to $\ket{0}_L$, which can be explained by the biased noise of spin qubits. However, this difference is limited, as the noise bias is mitigated by gate-induced errors, and the phase-flip to bit-flip conversion induced by Hadamard or equivalent ($K$) gates during syndrome extraction.

\subsection{P-gate Impact}\label{ap:P_gate_impact}
\begin{figure}[!t]
    \centering
    \includegraphics[width=\linewidth]{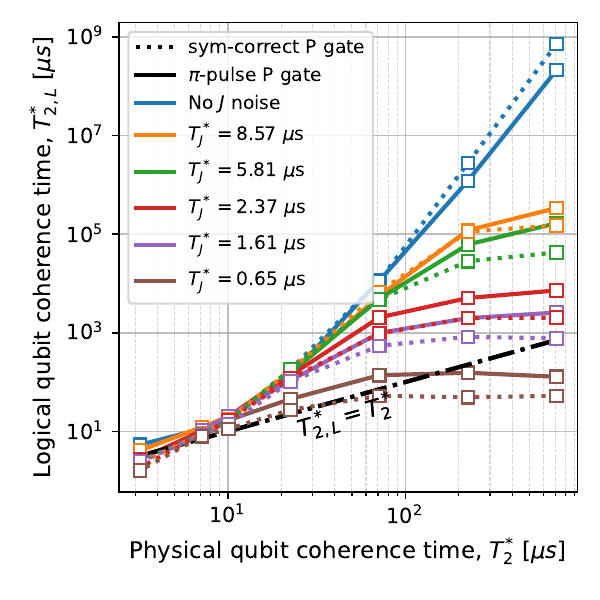}
    \caption{Logical qubit coherence time $T_{2,L}^*$ as a function of $T_2^*$, for different $T_J^*$ values induced by exchange energy noise, with syndrome measurement circuit using either the symmetry-corrected or the $\pi$-pulse $P$ gate.} 
    \label{fig:P_type_impact}
\end{figure}

Fig.~\ref{fig:P_type_impact} shows the QEC performance (in terms of logical $\TL$ vs. physical $T_2^*$), for two distinct syndrome measurement circuits: one using the symmetry-corrected $P$ gate, and the other one using the $\pi$-pulse $P$ gate. Even though the $\pi$-pulse $P$ gate is approximately six times longer than the symmetry-corrected $P$ gate ($t_{\text{sym-corr}} \approx 0.5$\,$\mu$s and $t_{\text{$\pi$-pulse}} = 3$\,$\mu$s), its enhanced resilience to both $\wqbit$ and $J$ noise---enabled by the spin-refocusing $\pi$-pulse sequence---leads to improved performance when $J$ noise is present. However, in the absence of $J$ noise, the symmetry-corrected $P$ gate induce comparable logical qubit performance, due to shorter gate duration.

\subsection{Rotated Surface Code vs. XZZX Variant}\label{ap:code_impact}

\begin{figure}[!t]
    \centering
    \includegraphics[width=\linewidth]{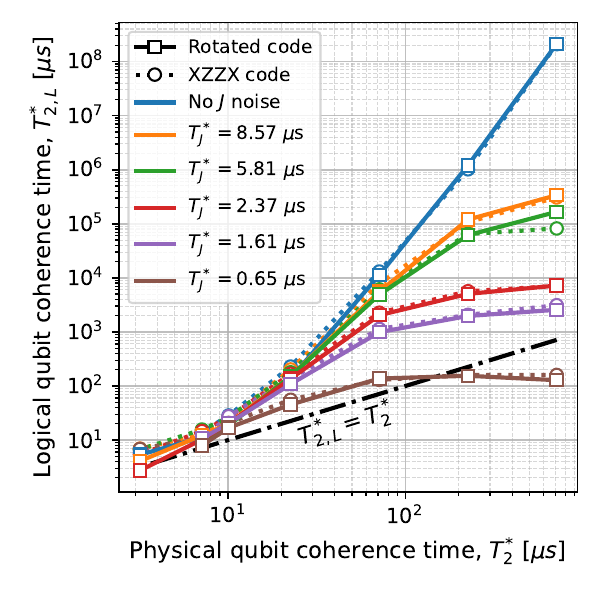}
    \caption{Logical qubit coherence time $T_{2,L}^*$ as a function of $T_2^*$, for different $T_J^*$ values induced by exchange energy noise, for  the 17-qubit rotated surface code and its XZZX variant. Both codes use the $\pi$-pulse $P$ gate  for syndrome measurements, and they exhibit similar performance.}
    \label{fig:code_impact}
\end{figure}

Fig.~\ref{fig:code_impact} shows the  logical $\TL$ vs. physical $T_2^*$, for both  the 17-qubit rotated surface code and its XZZX variant, known to be more robust for qubits experiencing biased noise~\cite{bonilla_ataides_xzzx_2021}. Our results show that both codes yield similar performance, due to the fact that the noise bias is mitigated by the syndrome measurement circuit (that is, gate-induced errors, and the phase-flip to bit-flip conversion induced by Hadamard or equivalent $K$ gates, as mentioned in Section~\ref{ap:log_type_compare}). 

\subsection{Quartic Dependence on $T_J^*$}\label{ap:TJ_dependance}
\begin{figure}[!h]
    \centering
    \includegraphics[width=\linewidth]{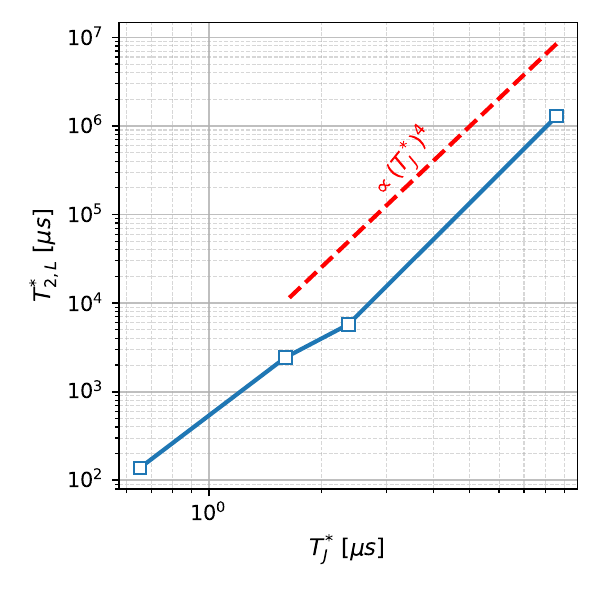}
    \caption{Logical qubit coherence time, $T_{2,L}^*$ as a function of $T_J^*$ for a fixed physical qubit coherence time of $T_2^* = 221.6\,\mu$s. As indicated by the red dashed line, $T_{2,L}^*$ follows a quartic dependence on $T_J^*$.}
    \label{fig:TJ_scaling}
\end{figure}

Figure~\ref{fig:TJ_scaling} presents the logical qubit coherence time as a function of $T_J^*$ for a fixed value of $T_2^* = 221.6\,\mu$s. A quartic dependence at high $T_J^*$ values is observed in this case, consistent with results shown in Section~\ref{sec:QEC_perfs}.

\newpage

\bibliographystyle{apsrev4-2} 
\bibliography{V9_article_bibli}
\end{document}